%% file: multi.tex
\documentclass[12pt,preprint]{aastex}

\usepackage{lscape}
\usepackage{soul}
\usepackage{datetime}

\newcommand{\gcmc}{\ensuremath{\rm g\,cm^{-3}}}
\newcommand{\teff}{\ensuremath{T_{\rm eff}}}
\newcommand{\teffp}{\ensuremath{T_{{\rm eff}p}}}
\newcommand{\logg}{\ensuremath{\log{g}}}

\newcommand{\feh}{[Fe/H]}

\newcommand{\rsun}{\ensuremath{R_\sun}}
\newcommand{\msun}{\ensuremath{M_\sun}}

\newcommand{\rstar}{\ensuremath{R_\star}}
\newcommand{\mstar}{\ensuremath{M_\star}}

\newcommand{\rhostar}{\ensuremath{\rho_\star}}
\newcommand{\rhoc}{\ensuremath{\rho_c}}

\newcommand{\rpl}{\ensuremath{R_{\rm p}}}

\newcommand{\teq}{\ensuremath{T_{\rm eq}}}
\newcommand{\rprs}{\ensuremath{R_{\rm p}/R_\star}}

\newcommand{\drho}{\ensuremath{d\rho_{c,i-j}}}
\newcommand{\adrs}{\ensuremath{a/R_\star}}

\newcommand{\rjup}{\ensuremath{R_{\rm J}}}
\newcommand{\rearth}{\ensuremath{R_{\earth}}}
\newcommand{\mjup}{\ensuremath{M_{\rm J}}}

\newcommand{\ikt}{{\it Kepler}}
\newcommand{\ik}{{\it Kepler~}}

\shorttitle{Hey look, we found lots of exoplanets.}
\shortauthors{Rowe et al.}
\slugcomment{Revision 8.0 --- \today --- \currenttime}

\begin{document}
%Hey look, we found lots of exoplanets.
\title{Validation of Kepler's Multiple Planet Candidates. III: Light Curve Analysis \&
Announcement of Hundreds of New Multi-planet Systems}

\author{Jason~F. Rowe,\altaffilmark{1,2} Stephen T. Bryson,\altaffilmark{1} Geoffrey W. Marcy,\altaffilmark{3} Jack J. Lissauer,\altaffilmark{1} \\
Daniel Jontof-Hutter,\altaffilmark{1,4} Fergal Mullally,\altaffilmark{1,2} Ronald L. Gilliland,\altaffilmark{5} Howard Issacson,\altaffilmark{3} \\
Eric Ford,\altaffilmark{6} Steve B. Howell,\altaffilmark{1} William J. Borucki,\altaffilmark{1} Michael Haas\altaffilmark{1}, Daniel Huber,\altaffilmark{1,4} \\
Jason H. Steffen,\altaffilmark{7,8} Susan E. Thompson,\altaffilmark{1,2} Elisa Quintana,\altaffilmark{1,2} Thomas Barclay,\altaffilmark{1,9} \\
Martin Still,\altaffilmark{1,9}  Jonathan Fortney,\altaffilmark{10} T. N. Gautier III,\altaffilmark{11}  Roger Hunter,\altaffilmark{1} \\
Douglas A. Caldwell,\altaffilmark{1,2} David R. Ciardi\altaffilmark{12} Edna Devore,\altaffilmark{2} 
William Cochran,\altaffilmark{13} \\
Jon Jenkins,\altaffilmark{1,2} Eric Agol,\altaffilmark{14} Joshua A. Carter,\altaffilmark{15} John Geary,\altaffilmark{15} \\
\email{Jason.Rowe@nasa.gov}
}

\altaffiltext{1}{NASA Ames Research Center, Moffett Field, CA 94035}
\altaffiltext{2}{SETI Institute, Mountain View, CA 94043}
\altaffiltext{3}{University of California, Berkeley, CA, 94720}
\altaffiltext{4}{NASA Postdoctoral Program}
\altaffiltext{5}{Center for Exoplanets and Habitable Worlds, The Pennsylvania State University, University Park, PA, 16802, USA}
\altaffiltext{6}{Pennsylvania State University, PA 16801}
\altaffiltext{7}{Northwestern University, Department of Physics \& Astronomy/CIERA, 2145 Sheridan Road, Evanston, IL 60208}
\altaffiltext{8}{Lindheimer Fellow}
\altaffiltext{9}{Bay Area Environmental Research Inst., 596 1st St West, Sonoma, CA 95476}
\altaffiltext{10}{Department of Astronomy and Astrophysics, University of California, Santa Cruz, CA 95064}
\altaffiltext{11}{Jet Propulsion Laboratory, California Institute of Technology}
\altaffiltext{12}{NASA Exoplanet Science Institute/Caltech}
\altaffiltext{13}{Department of Astronomy and McDonald Observatory, The University of Texas at Austin}
\altaffiltext{14}{Department of Astronomy, Box 351580, University of Washington, Seattle, WA 98195, USA }
\altaffiltext{15}{Harvard-Smithsonian Center for Astrophysics, 60 Garden St., Cambridge, MA 02138}

\begin{abstract}

The \ik mission has discovered over 2500 exoplanet candidates in the first two years of spacecraft data, with approximately 40\% of them in candidate multi-planet systems.  The high rate of multiplicity combined with the low rate of identified false-positives indicates that the multiplanet systems contain very few false-positive signals due to other systems not gravitationally bound to the target star (Lissauer, J.~J., et al., 2012, ApJ 750, 131).   False positives in the multi-planet systems are identified and removed, leaving behind a residual population of candidate multi-planet transiting systems expected to have a false-positive rate less than 1\%.  We present a sample of 340 planetary systems that contain 851 planets that are validated to substantially better than the 99\% confidence level; the vast majority of these have not been previously verified as planets.  We expect $\sim$2 unidentified false-positives making our sample of planet very reliable.  We present fundamental planetary properties of our sample based on a comprehensive analysis of \ik light curves and ground-based spectroscopy and high-resolution imaging.  Since we do not require spectroscopy or high-resolution imaging for validation, some of our derived parameters for a planetary system may be systematically incorrect due to dilution from light due to additional stars in the photometric aperture.  None the less, our result nearly doubles the number of verified exoplanets.

%The \ik mission is the best thing since sliced bread.

\end{abstract}

\keywords{planets: Kepler-100, Kepler-101,..., Kepler-421 --- Facilities: \facility{The \ik Mission}.}

\section{Introduction}\label{intro}

%To Do List:
%\begin{itemize}
%\item hand fix period for 490.02 to 800d
%\item remember to remove -- from multiprops.dat before run stats prog
%\item fits for KOI377.01 and .02 seem wrong (check that TTVs are used properly)
%\item add footnotes to previously validated systems. 0-not validated by multiplicity, 1-we would of validated via multiplicity
%\item Steve B. does not go back to look at a planet if it was previously validated.  Should be noted in text
%\item Jillo-Box Speckle paper - make Steve aware of this 
%\item all the figures that show the wonderful transits
%\item determine how many of the {\bf validated} multis hosts have KECK spectra
%\item check auto-correlation for MCMC
%\end{itemize}

Data from the first two years of \ik spacecraft operations have identified 3670 target stars with periodic or transit-like signatures indicative of transiting planets or eclipsing binary stars.  Approximately 50\% of these targets have signatures that can be attributed to false-positives (FPs), primarily eclipsing binaries (EBs) centered on the target star, a chance alignment of a distant EB within the photometric aperture, or flux bleeds into the photometric aperture.  The remaining 2530 systems are composed of primarily exoplanetary systems with an expected FP rate of approximately 10\% due to photometric blends \citep{mor11,fre13,san13}.  However,  a subset of 457 systems show more than one candidate transiting planet candidate (PC); we refer to these candidates as ``multis". FPs should be nearly randomly distributed among \ik targets, whereas if flat multi-planet systems are common, then many targets should have multiple transiting planets \citep{lis11}.  The large number of multis observed thus implies a high reliability rate, as quantified by (\citealt{lis12,lis13}; henceforth Paper I and Paper II, respectively).  The small number of FPs found in multis observed by \ik \citep[and \S\ref{vps} of this paper]{lat11} reinforces our confidence in the high reliability of the PCs remaining in multis (see Appendix C of Paper II for details). After making selections to minimize the odds of blend scenarios, we find 340 systems containing a total of 851 planets that can be validated to better than the 99 percentile, with 768 planets across 306 systems being newly validated.  Some of these systems have also been confirmed from radial velocity detection \citep{mar13,gau12}, transit timing variations \citep{for11,ste12} and planet validation techniques such as {\it BLENDER} \citep{tor11,fre12} and now through multiplicity boost (Paper I).  We increase the known number of exoplanets from 942\footnote{Based on NASA Exoplanet Archive 2013/11/12} to 1710.   

With excellent precision and high-duty cycle, \ik observations of transiting exoplanet systems provide photometric data that can be used to measure fundamental transit properties such as transit duration and observed transit depth.  These properties are determined by the relative sizes of the planet, host star and additional flux sources (other stars) within the photometric aperture.  Since observations provide nearly full coverage of the planetary orbit, the resulting photometric phase curves enable a useful diagnostic for the identification of astrophysical FPs that can mimic an exoplanet transit signature.  In \S \ref{lca} we examine the light curves of the \ik sample and describe the nature of the planetary systems and how they are identified.

As an imaging instrument, \ik also provides time-series measurements of the centroid of the photometric signal.  When multiple sources are present in the photometric aperture, the photometric centroid can move in response to flux changes from any of the sources.  This property allows \ik pixel level data to be used to search for scenarios where a planetary transit-like event is produced by a diluted background eclipsing binary star \citep{bat10}.  

Paper II presents a theoretical exploration of the expected and predicted FP rate for transiting multi-planet systems.    From the $\sim$190 000 targets observed by \ikt, there are roughly 2500 transit-like patterns of events in the KOI catalogue, split between PCs and EBs.  Thus, for a \ik target chosen at random it is unlikely that a transit event will be present.  One of the most common classes of FP event is caused by a background eclipsing binary (BEB) in the aperture of the target star.  The source of a BEB can either be an additional star that is found within or close to the photometric aperture or a bright star within the \ik Field of View (FOV) that introduces flux in the photometric aperture due to optical ghosting, such as mirror images, or electronic interference, such as CCD crosstalk \citep{cou13}.   The occurrence rate of FPs is largely independent of the target stars and thus it is far more likely that a FP source will produce a single transit-like event as opposed to a photometric light curve containing transit-like signals from multi-planet sources.   It is important to distinguish between FPs produced by background stars from those caused by instrumental effects.   We identify the latter as period/phase (P/T0) collisions to indicate that the period and epoch solution for the candidate event are not due to a unique event.  Rather, the transit signal from one target is seen on another target as well.  P/T0 collisions account for most of the identified instrumental FPs.  A full description and catalogue of P/T0 collisions can be be found in \citet{cou13}.  Figure 1 of Paper II shows the galactic distribution of \ik targets, PCs and FPs.  The results indicate that only a small fraction of the remaining \ik PCs are likely to be background eclipsing binary (BEB) FPs. 

The process of identifying and cataloguing FPs from the KOI list continues to evolve.  Thus, the FP list from various iterations of the KOI catalogue \citep{bor11,bat12,bat13} have different reliability rates.  This paper describes in detail the steps taken to develop a reliable and uniform classification scheme and its application to the multi-planet sample to classify KOIs as false-alarms (FAs), FPs or PCs.  FAs are transit event candidates with a signal-to-noise ratio (S/N) below a S/N threshold of 7.1 (see \S \ref{snr}) or a transit candidate mimicked by stellar variability or an instrumental artifact.  We find that the FP rate for multi-planet systems is low and consistent with the predictions from Appendix C of Paper II based on a statistical analysis of the FP rate found in the single planet population.  The predicted FP rate allows us to conclude that the PCs that pass our FP and morphological tests would misclassify only $\sim$2 PCs, allowing us to claim that 768 PCs are bona fide planets with a confidence level greater than 99\%.

As an example, consider only the non-transiting, transiting single planet and transiting double planet systems.  Here we ignore the case of FP+FP, where two FPs are associated with the same target and estimate the number of FP for the double planet systems in the spirit of Paper II.  There are approximately 140 000 stars that contribute to the transiting planet population (\S 4 of Paper II).  After removing FAs and T/P0 collisions there are 2182 systems that show one transiting body and 284 systems that show two transiting bodies.  From the single-planet systems, 662 systems were identified as FPs, which provides an estimate of the FP rate of 0.44\%.  Thus, from the 1500 good single-planet candidates 7 PC+FP systems are predicted.  From the sample of 295 that have two transiting candidates, there are 6 that were identified as a PC+FP combination.  Good agreement is found even in our simplified case.  Of course, one needs to properly account for EBs and the entire range of multi-planet candidates and the multitude of PC and FP mixes that can be produced. These considerations are the basis of Papers I and II, which provide predictions of FPs rates that are verified in this paper.

The combined analysis reported in Paper II and this manuscript validates more than 300 new Kepler multi-planet systems.  Paper II introduces the binary star planet hosts Kepler-132, where one star hosts two transiting planets and its companion hosts one transiting planet and Kepler-296, a pair of small stars with a total of 5 transiting planets, the multi-resonant 4-planet Kepler-223 system and two additional planets in the Kepler-80 = KOI-500 system that includes two 3-body resonances, as well as several high-multiplicity systems, including the new 5-planet systems Kepler-102 = KOI-82, Kepler-169 = KOI-505, Kepler-238 = KOI-834 and Kepler-292 = KOI-1364, three new planets orbiting Kepler-84 = KOI-1589 (bringing the total count to 5) and partial validations of the 5-candidate systems Kepler-122 = KOI-232 (4 planets validated) and Kepler-154 = KOI-435  (2 planets validated).  Hundreds of new planetary systems are announced herein, with special attention given to four new planets with radii roughly twice that of Earth located in or near the habitable zones of their host stars.

This paper is organized as follows: In \S\ref{sample} a description of the planetary sample is presented.   The adopted stellar parameters are discussed in \S\ref{stellarpars}.  Detailed descriptions of the transit models and lightcurve analysis are described in \S\ref{lca}.   The process of identifying FPs can be found in \S\ref{fpi}. Subsection \ref{centroids} covers centroid measurements that provide a clean sample of highly probable transiting multi-planet systems which we demonstrate in \S\ref{vps} are genuine extrasolar planets.   We conclude with a discussion of the multi-planet population in \S\ref{pop}.
%Hierarchical blends and the eccentricity distribution of the multi-planet sample are explored in \S\ref{tdur}.

\section{Planet Candidate Sample}\label{sample}

Photometric surveys for extrasolar planets are contaminated by FPs that are caused by eclipsing stellar binaries and transits and eclipses of stars that are spatially offset from the target stars.   The Kepler-Object-of-Interest (KOI) catalogue is an inhomogeneous working list used to track transit candidates of interest identified from {\it Kepler} photometric light curves.   The FP to PC ratio of the raw KOI catalogue is approximately 0.39 \citep{bur13}\footnote{Based on KOIs 1-3149 having a FP or PC status on 2013/08/01}.  A quick survey of KOI dispositions (available on the \ik exoplanet archive; \citealt{ake13}) shows most of the FP occurrences are linked to events that show a single pattern of periodic transits.  They rarely occur when there is evidence of multiple transiting objects, a fact first noted by \citet{lat11}.  While the vast majority of the catalogue is dedicated to exoplanet candidates and eclipsing binary stars, the list contains some astrophysically interesting light curves that do not show transits, such as {\it Heartbeat Stars} \citep{tho12}.  We have excluded events that have been classified as a {\it non-transit event}.  Similarly, false-alarms (FAs) were also excluded based on a transit S/N of 7.1. The \ik transit detection pipeline was designed to identify events that have at least 3 transits.  However, through human examination of the photometric time series, some deep, single occurrence events are noted and catalogued.   The orbital periods for some of these events have now been measured based on the detection of a third transit.  However, there are 18 planet candidates for which only 1 or 2 transits have been observed.  It is difficult to assess the FP nature of these candidates as one cannot reliably estimate the orbital period. These KOI are also excluded from our sample.

Given that the \ik pipeline is continually undergoing substantial improvements, we restrict ourselves to exoplanet candidates found with Q1 -- Q8 light curves. This includes the Q1 -- Q5 \citep{bor11} and Q1 -- Q6 \citep{bat13} lists and a subset of the Q1 -- Q8 catalogue \citep{bur13}. The Q1 -- Q8 candidate list includes data products based on Q1 -- Q10 data. During this process additional multi-planet candidates were discovered and catalogued, including planet candidates found with independent methods such as {\it QATS} \citep{car13}.   An example is KOI-351 (Kepler-90) which had 6 PCs found in the Q1 -- Q8 sample but is now known to contain at least 7 candidates (Paper II and \citealt{ago13}).  We have excluded these Q1 -- Q10 discoveries to avoid a bias in our statistics that would overestimate the quantity of multi-planet systems for validation.  Any multi-planet candidate found in Burke et al. (2013) that is not listed in our sample is a Q1-- Q10 discovery. We used 3737 KOIs associated with 3008 stars in our analysis; 1210 planet candidates in multi-planet systems and 2527 planet candidates in single planet systems. 

Table \ref{multicount} lists the number of systems and planets that have been considered in our sample.  Each row lists the number of systems that pass various tests, such as our FA test (\S\ref{snr}), P/T0 collisions and transit characterization from models (\S\ref{mcmc}). The columns indicate the number of planet candidates found in each system with S1 indicating that one planet candidate was found.  M1 means that a light curve was identified to have multiple transiting candidates, but after cuts the number of remaining candidates has fallen to one.  The columns labeled 2 through 6 indicate the number of systems with the corresponding number of planets that pass criteria indicated in the last column.  Each test was chosen to eliminate and mitigate FPs (see \S3 of Paper II), but only candidates labeled as FP are definitive FPs.  For example, we do not validate candidates that show `V'-shaped light curves based on transit model fits.  This distinctive transit shape can be produced by either a transit duration that is similar to the photometric cadence, a grazing planet with an impact parameter near 1, or stellar binary with a large value of \rprs.  Transit-modelling of a `V'-shaped transit leads to a large uncertainty on the measurement of quantities such as \rprs, making it difficult to assess the properties of the planet-candidate and identify blended BEBs.  These candidates are not labeled as FPs, but they are also not validated as planets. 

Approximately one-third of the entries in the KOI catalogue have been identified as eclipsing binaries either through light curve or centroid analysis.  Criteria for KOI promotion have not been uniformly applied, and there have been some systematic biases in categorization based upon multiplicity of sets of transit signatures. For example, a low amplitude sequence of dips with alternating depths is generally categorized as an eclipsing binary and not given a KOI number, but if such a signature is seen for a target that has already has been classified as a KOI based upon another set of dips, it is given a KOI number and immediately classified as an FP, thereby providing a biased increase in the number of FPs in multis. Searching for additional candidates is terminated around targets that have been identified as FPs, although there is often a significant amount of time between initial identification as a KOI and labeling as a FP. If a target is first identified as an EB, then the pipeline transit search is terminated, and since most such identifications are done rapidly, the distribution is more strongly biased against combinations of EBs and planet candidates.  To improve our estimate of the true FP rate, we supplement our FP list with the eclipsing binary catalogue\footnote{V3 retrieved 2013/04/24 http://keplerebs.villanova.edu/}.  Since we are interested in EBs that roughly match the signal from a transiting extrasolar planet we only considered EBs that are detached based on having a morphological classification criteria less than 0.5 \citep{mat12} and have a primary eclipse depth less than 2\%.  These criteria remove contact binaries, which have a characteristic {\it w} shape. This step increases the total number of FPs by 138.

%However, it is usual to exclude obvious EBs from the KOI catalogue.  For example, candidates that show strong ($>$ 0.1\%) occultations due to the transiting object passing behind the primary star are almost certaintly EBs. Nevertheless some of these candidates were assigned a KOI identifier and others were not.  

\section{Characteristics of Planet-Hosting Stars}\label{stellarpars}

We used a diverse set of measurements to  estimate the properties of each stellar host of the \ik multis that we validate as planets.  Our goal is to obtain the best classification of each planet-hosting star given all of the information available to us rather than to produce a homogeneous data set. We prioritized classification in the following order, choosing for each target the  first available option:

\begin{itemize}

\item Combined asteroseismology + spectroscopy analysis  \citep{hub13}.  

\item Spectrometry Made Easy (SME; \citealt{val96}) analysis using spectra taken at the Keck I telescope.

\item SpecMatch fitting (see below and \citealt{pet13}) using spectra taken at the Keck I telescope. 

\item Stellar Parameter Classification (SPC) analysis of spectra taken at various telescopes \citep{buc12}. 

\item Modified KIC photometric classification from adjustments to original KIC values of $T_{\rm eff}, \log{g}$ and [Fe/H] to match Yale-Yonsei \citep{dem04} stellar evolution models.  
%Specific details of the procedure can be found in Section 3 of Paper III.
\end{itemize}

\noindent Note that by using this heterogeneous set of stellar characterization techniques, we sacrifice uniformity for accuracy, and care should be taken in performing statistical studies based on fit parameters.

To support the SME and SpecMatch analyses, high resolution spectra were taken of multi-planet candidate host stars with HIRES spectrometer on the Keck I telescope using the observing setup of the CPS group \citep{mar08}. We acquired spectra with a resolution of R = 55,000 and a wavelength coverage of 360 -- 800 nm, which have a S/N per pixel of 40 (or better in some cases that were used for the SME analysis) at 550 nm, corresponding to a S/N = 85 per resolution element. The spectra were observed without the iodine cell in the light path.  Using the C2 decker, which projects to $0{\farcs}87 \times 14{\farcs}0$ on the sky, we removed the signal from moonlight that otherwise could contaminate the stellar spectra at the level of a few percent.

When determining atmospheric parameters of the planet host stars using SpecMatch, we compared each spectrum to a library of 800 spectra having $T_{\rm eff} = 3500-7500$ K and $\log g  = 2.0 - 5.0$, which spans the FGK and early M type main sequence and subgiant stars. All library stars have accurate parallax measurements, allowing for good estimates of stellar mass and radius for each. We then compared the observed spectrum with that of each library star. The spectrum is placed on a common wavelength scale and normalized in intensity. The $\chi^2$ value is calculated as the sum of the squares of the differences between the observed spectrum and each library spectrum. The final stellar properties, listed in Table \ref{spars}, are determined by the weighted mean of the ten library spectra with the lowest $\chi^2$ values. We adopted errors in each parameter by comparing results to a range of standard stars.

Stellar parameters are derived by matching atmospheric parameters (\teff, \logg, \feh) to stellar evolution models (\mstar, Age, Z).  Atmospheric parameters are based on SME \citep{val96}, SpecMatch, SPC \citep{buc12}, Asteroseisomology \citep{hub13} or the KIC \citep{bro11} including the revision of \teff\ by \citet{pin12}.   For SME parameters we added 59 K to \teff\ and 0.062 dex to \feh\ in quadrature following the \citet{tor12} recommendation.  For SpecMatch and SPC results we adopt uncertainties as reported, with preference given to SME and then SpecMatch.  For stars without asteroseismic or spectroscopic constraints, we adopt \teff\ from Table 7 of \citet{pin12} and \logg\ and \feh\ as given in the KIC.  For uncertainties we adopt values of 200 K in \teff, 0.3 dex in \logg\ and 0.5 dex in \feh, in agreement with typical residuals of KIC values to stellar properties determined from asteroseismology and spectroscopy (e.g., \citealt{bru12}).  We adopted the Yonsei-Yale stellar evolution models \citep{dem04} to determine stellar parameters.  The model matching was done by varying the stellar mass, age and Z and comparing the model-derived values of \teff, \logg\ and \feh\ with the spectroscopic values with a chi-square statistic.  An initial match was found by scanning in mass increments of 0.1 \msun\, restricting ages from 0 to 14 Gyr, and identifying a best matching model.  A Markov-Chain-Monte-Carlo (MCMC) routine was then seeded with this trial value of stellar mass, age and Z to determine posterior distributions.  All stellar models with ages greater than 14 Gyr were excluded.  In total 100 000 chain elements were generated for each star.  The models were also used to determine posterior distributions for the stellar radius, luminosity and mean stellar density.  The resulting stellar parameters are listed in Table \ref{spars}.

%Stellar parameters for higher mass stars are tricky because of degeneracies in Teff-\logg-\feh\ space.
%These are the stellar parameters found on nexsci.

\section{Light Curve Analysis}\label{lca}

\ik photometry was used to both identify FPs and to characterize the transiting planets. We used \ik Q1 to Q10, long cadence, simple aperture, photometric observations\footnote{Observations labeled as SAP\_FLUX from FITS files retrieved from The Barbara A. Mikulski Archive for Space Telescopes (MAST).} gathered every 29.4 minutes over a time span of 868 days.  These measurements do not account for the effects of dilution from the addition of stars near or in the photometric aperture, thus, there is a bias in our measured planetary parameters towards underestimating planetary radii.  Measuring dilution and determining corrections are difficult tasks and outside the scope of our goal to validate hundreds of extrasolar planets.  However, there are rough estimates of the dilution based on the KIC, retrieved from MAST, from which this bias can be estimated and used to place conservative upper limits on systematics introduced from contamination.  The mean value of light contamination for validated \ik planets is 5\%.  As the transit depth is proportional to $(\rprs)^2$, a 5\% dilution translates into a 2.6\% systematic bias in the planetary radius, which is small compared to the uncertainty in the stellar radius.  From our validated planet sample, the largest light contamination was found to be 20\% of the total light for KOI-907 (Kepler-251), which translates into an error on the planetary radii of 11.8\%.

While \ik had a high duty cycle,  some  transits were missed due to a variety of logistical details such as sky location, data downlink, spacecraft safe modes and a dead module.  An extreme example is KOI-94 (Kepler-89) \citep{wei13}, which has an effective duty cycle of less than 50\% due to its location on the edge of one of the CCD detectors.  The spacecraft rotates each quarter and this target lies in the gap between detectors during 2 of every 4 consecutive quarters.  For a majority of our targets, the effective duty fraction after removal of flagged\footnote{Observations with SAP\_QUALITY=0 from FITS files retrieved from MAST} data was approximately 86\%. 

We filtered the data to remove instrumental and astrophysical signatures that are independent of the planetary transit as follows: each observation was corrected by fitting a cubic polynomial to a segment of the time series photometric measurements centered on the time of measurement.  A segment is defined by selecting observations that were taken within 1 day of the measurement.  We also require that the time series not contain any gaps longer than 5 cadences ($\sim$ 2.5 hours).  If such a gap is encountered, the data collected near that gap are not considered.  Such gaps were commonly produced by the monthly data downlinks.  The removed data dropped the duty-cycle by $\sim$1\%.  After repointing the spacecraft, there was usually a photometric offset produced due to thermal changes in the telescope.  Thus, astrophysical signals with timescales of approximately 2 days are strongly filtered by this process.  The filter is destructive to the shape of a planetary transit.  Thus, we exclude any measurement taken within 1 transit-duration of the measured center of the transit time and use an extrapolation of the polynomial fit to estimate corrections during transits.  The transit duration is defined as the time from first to last contact, ${\rm T}_{dur}$.  The segment is fit with a cubic polynomial and used to measure the photometric offset, which is then removed.  We repeat the process for each observation to produce a detrended time series.  When significant transit timing variations are detected, we rebuild the detrended time series using the updated center of transit times.

An initial multi-planet photometric model was fit to each detrended light curve.  The photometric model assumes non-interacting circular orbits and used the quadratic limb darkening transit model of \citet{man02}.  We used limb-darkening parameters from \citet{cla11}, which were fixed for each target based on our stellar classification (\teff, \logg, \feh).  The model was parameterized by the mean-stellar density (\rhostar), photometric zero point and for each planet ($n$) an epoch ($T0_n$), period ($P_n$), scaled planetary radius (\rprs$_n$) and impact parameter ($b_n$).  The semi-major axis for each planet candidate is estimated by
\begin{equation}\label{rhostar}
\left( \frac{a}{\rstar} \right)^3 \simeq \frac{\rhostar G P^2}{3 \pi},
\end{equation} 
where the assumption was made that the sum of the planetary masses is much less than the mass of the host star.  For a Jupiter-mass companion of a Sun-like star, a systematic error of 0.1\% is incurred on the determination of \rhostar.  To account for the $\sim$30 minute integration of \ik observations, the transit model was sampled 11 times temporally with equal spacings within the integration window.  The 11 separate models were then averaged.   A best fit model was calculated by a Levenberg-Marquardt chi-square minimization routine \citep{mor80}.  This model was used primarily to seed our Markov Chain Monte Carlo MCMC) routines to measure fundamental physical properties of each planet.

\subsection{Measuring Planet Parameters}\label{mcmc}

Our main objective is to identify FPs and to select candidates found in multi-transiting systems that have a very high probability of being bona fide extrasolar planets.  Our strategy was to examine each photometric light curve for signatures of stellar binarity: secondary eclipses, phase curve variations and a comparison of the transit model determination of \rhostar\ to our classification and modeling of the host star.  We also examine the populations of stellar and planetary systems to establish regions of parameter space, namely orbital period and impact parameter, that are most susceptible to contamination from FPs.

Our measured planetary parameters are listed in Table \ref{mplanetfit} and are based on a transit model fit similar to the description given at the beginning of \S\ref{lca}, except that we have modeled each planet candidate in a system independently.  We start by using the best fit model from the multi-planet model to remove the photometric signature of all transiting candidates except the one we wish to measure.  We assumed a circular orbit and fit for \rhostar, T0, $P$, $b$, \rprs\ and \rhoc, where \rhoc\ is the value of \rhostar\ when a circular orbit is assumed.  Thus, each planet candidate provides an independent measurement of \rhoc.  If the value of \rhoc\ is statistically the same for each planet candidate, then the planetary system is consistent with each planet being in a circular orbit around the same host star.  We examine the distribution of transit-determined values of \rhoc\ in Appendix \ref{tdur}.

To estimate the posterior distribution on each fitted parameter, we use a MCMC approach similar to the procedure outlined in \citet{for05}.  To account for the strong correlation between \rhoc, $b$ and \rprs, we use a Gibbs sampler to shuffle the value of parameters for each step of the MCMC procedure and use a control set of parameters to approximate the scale and orientation for the jumping distribution of correlated parameters as outlined in \citet{gre11}.  An initial control set consists of 2000 chains was generated by a MCMC run where the width of the Gaussian proposal distributions was adjusted to achieve a success rate of $\sim$25\%.  Once the success rate for a jump was between 20 and 30\%, the width of the Gaussian was fixed for the duration of the calculations.  The control set is updated during the MCMC run by adding every second accepted jump proposal parameter set and removing the oldest element.  The control set is fixed once an acceptance rate between 20-25\% is achieved.  Any chain that was generated before the proposal sample was fixed was discarded.  We found that this method allows the MCMC approach to efficiently sample parameter space even with highly correlated model parameters.   We generated four 1 000 000 Markov-chains for each PC. The first 20\% of each chain was discarded and the remaining sets were combined and used to calculate the median, standard deviation and $1,2,3\sigma$ bounds of the distribution centred on the median of each modeled parameter.  Our model fits and uncertainties are reported in Table \ref{mplanetfit}.  We use the Markov-Chains to derive model dependent measurements of the transit depth (T$_{dep}$) and transit duration (T$_{dur}$).  We also convolve the transit model parameters with the stellar parameters (see \S\ref{stellarpars}) to compute the planetary radius, \rpl, and the flux received by the planet relative to the Earth ($S$).  To compute the transit duration, we used Equation 3 from \citet{sea03} for a circular orbit,
\begin{equation}
{\rm T}_{dur} = \frac{P}{\pi} \arcsin \left(  \frac{\rstar}{a} \left[ \frac{(1+\frac{\rpl}{\rstar})^2-(\frac{a}{\rstar}\cos i)^2}{1-\cos^2i} \right]^{1/2}  \right),
\end{equation}
which defines the transit duration as the time from first to last contact.  We estimate the ratio of incident flux received by the planet relative to the Earth's incident flux,
\begin{equation}
S = \left( \frac{\rstar}{\rsun} \right)^2 \left( \frac{\teff}{T_{{\rm eff}\sun}} \right)^4 \left( \frac{a}{a_{\earth}} \right)^{-2},
\end{equation}
where \teff\ is the effective temperature of host star, $T_{{\rm eff}\sun}$ is the temperature of the Sun and $a$ is the semi-major axis calculated with Kepler's Second Law using the measured orbital period and estimated stellar mass.  
%We list all our measured planetary parameters in Table \ref{mplanetfit}.

\subsection{Transit Timing Variations}\label{ttv}

We estimate transit timing variations (TTVs) for each light curve using the best fit models from \S\ref{lca} and \S\ref{mcmc} as a template.  Center-of-transit times are measured by selecting data obtained within 1 transit duration of the predicted center of transit time (thus the time series has a length that is twice the transit duration).  If the transit duration is less than 2 hours, then we select data within 2 hours of the center-of-transit time. We then refit the transit model but we allow only T0 to vary.   The measured center-of-transit time is then compared to the predicted time to produce the observed minus calculated transit time (TTV$_n$) for each transit, $n$.  If significant variations are detected, we improve the transit model template by compressing and expanding the time interval between measurements by linearly interpolating between timing offsets observed for each transit.  The improved template is then used to redetermine the transit times.  We report our measured transit-timing variations in Table \ref{ttvcat}.  If fewer than 4 observations were selected for fitting, we do not report TTV$_n$.  Note that here $t_n$ is the measured transit time and not the prediction of a linear transit ephemeris (unlike \citealt{for11}). 

While the vast majority of the TTVs were processed in bulk, some KOIs with large TTVs received individual attention.  When the center-of-transit time was shifted substantually away from the predicted transit time, the fitting process failed.  An example is KOI-142 (Kepler-88), where the transit times shift by $\sim$20 hours.  For such cases, the previous two transit timing measurements were used to linear extrapolate an estimate of the next transit time to initialize the fitter.

\section{Planet Dispositions and False Positive Identification}\label{fpi}

The adopted Q1--Q8 dispositions were produced by a combination of work developed for the general KOI catalogue \citep{bur13} and the multi-planet population listed in this paper.  The end result is a set of dispositions shared between the two papers.  Each planet candidate was subjected to tests described below: lightcurve inspection by eye, a S/N threshold, searches for secondary events, phase-linked variations, odd-even numbered transit comparison and centroid motion during transit.  The disposition of KOIs presented here and the underlying statistics presented in Paper II sets the stage for the validation of a large number of multi-planet candidates at greater than the 99 percentile; specifically, we expect $\sim$2 FPs from the 851 planets validated in \S\ref{vps}.

One of the major results from Paper II (see \S 4 therein) is that the FP rate in multi-planet systems must be low.  The predictions are that $\sim$27 FPs should have been detected in the multi-planet candidate sample, and that $\sim$2 FPs have been missed.  Demonstration of the accuracy of the first of these predictions builds a strong case that currently viable planet candidates in multi-planet systems are bona fide planets.  The types of FPs that can be searched for include: a planet transiting a star not physically bound to the \ik target star; or an eclipsing binary star system or other astrophysical phenomenon.  If a bound stellar companion is found, it is sometimes possible to determine which star is the source of the transits, but \ik data are not sensitive to isolating the transit host in most bound systems.

This section is dedicated to describing the tests that were carried out to identify FPs in our multi-planet sample.   These tests include searches for secondary events and classifying the event as a planetary occultation or stellar eclipse (\S\ref{occ}).    Tidal interactions and motion of the host star around the center-of-mass produce variations related to the orbital period.  When present, the amplitude of these variations reveals the masses of the binary components (\S\ref{phase}). In the cases where the primary and secondary eclipses of a stellar binary with nearly equal mass stars are reported, the orbital period may be incorrect by a factor of 2.  To test for this scenario we compare the depths of the odd- and even-numbered transit events (\S\ref{oddeven}).  One of the most powerful tests is the use of pixel-level Kepler data to focus on finding clean targets for validation by localizing the source of the transit on the detector (\S\ref{centroids}).   A common source of FPs are centroid offsets due to motion in the difference of in- and out-of-transit combined images across a transit event.  The most frequent source of centroid offsets are background eclipsing binaries that track the spatial density of background stars (see Figure 1 of paper II).

\subsection{Quality of Model Fit}

We calculated the reduced chi-square for each transit model.  If the value was greater than 2 or less than 0.5, then the fit and photometry were visually inspected.  In most cases, it was found that a transit overlapped an instrumental effect, the most common effect being photometric deviations observed after the instrument returned to nominal operations that involved a reorientation of the spacecraft.  In these cases, the offending segment of data was excluded and the model fits were repeated.  Other cases include models that produced a poor transit fit from convergence to a local minimum, excess scatter from stellar variability, and evidence of a stellar binary in the light curve shape from the presence of a strong occultation or ellipsoidal variations.  This level of vetting was performed for both the single and multi-planet population.   

From our inspection we discovered that KOI-1134.01 and 1134.02 were both tracing the same EB that had a period of 100 days but had transit depths that were heavily modulated due to third-light contamination that appeared as independent transit candidates in early analysis using only a few months of observations.  We have labeled KOI-1134.01 as an EB FP and 1134.02 was labeled as a FA.   KOI-1792.02 was found to be a FP with stellar variability mimicking a transit signal.  KOI-2048.02 was identified as residuals of the transit fit to KOI-2048.01, thus 2048.02 was labeled as a FA and removed.

\subsection{Signal-to-Noise Ratio}\label{snr}

The signal-to-noise ratio (S/N) was calculated from depth of the transit using the transit-model and noise was estimated by the standard deviation of observations obtained outside of transit and then scaled with a geometric sum to match the transit duration, yielding
\begin{equation}
{\rm S/N} = \sqrt{N_T} \frac{{\rm T}_{dep}}{\sigma_{OT}},
\end{equation}
where {\rm T}$_{dep}$ is the transit depth, $N_T$ is the number observations obtained during transit and $\sigma_{OT}$ is the standard deviation of out-of-transit observations.   The estimation of the S/N assumes that the depth of the transit is uniform, which is a good approximation for small Earth-sized planets with central transits ($b$=0).   For relatively large planet-to-star radius ratio and/or large impact parameters, our technique will overestimate the S/N, but this has minimal impact on our assessment of PCs.  The impact parameter is not well defined for low S/N events, but the transit depth and duration are measurable quantities.

We used S/N estimates for two purposes: to identify FAs and to determine a threshold for planet validation.  The KOI catalogue has an adopted S/N limit of 7.1 to classify a target as a KOI.  FAs are present in the KOI catalogue as initially the transit signal was estimated to have a S/N greater than 7.1 \citep{jen02}, and then as additional observations were gathered the S/N dropped below 7.1.  These types of events inform us that validation of transiting planets with a S/N near the KOI threshold has a risk of introducing FAs, which we will now assess.

Using a S/N cut of 7.1 based on the transit depth and visual inspection, 26 KOIs in multis were classified as FAs: KOI-111.04, 439.02, 489.02, 966.02, 989.01, 1070.03, 1134.02, 1198.04, 1312.01, 1316.02, 1408.02, 1576.03, 1639.01, 1639.02, 1792.02, 1940.02, 1961.02, 2048.02, 2160.02, 2188.02, 2224.02, 2261.02, 2339.02, 2473.02, 2533.02, 2586.02.  For 15 systems, only a single planet candidate remained, as indicated by the {\bf M1} column in Table \ref{multicount} when the S/N cut is applied.   The objects were considered single-planet systems for statistical counts.  The rate of FAs from the single-planet and multi-planet population were both found to be $\sim$2\%.  As FAs do not represent real detections (quite the opposite) there is no reason to expect the rates to be predictable or reliable.  The KOI creation process has been very inhomogeneous. This has caused the introduction of biases that favour finding and identifying additional planet candidates once the first candidate in a system has been found. This is especially true because of the notion that the FP rate for multi-planet systems is low.  Quantifying this human bias is difficult and is part of our motivation to choose a larger transit S/N cut of 10 for planet validation.

\begin{figure}
\begin{center}
\includegraphics[scale=0.5,angle=270]{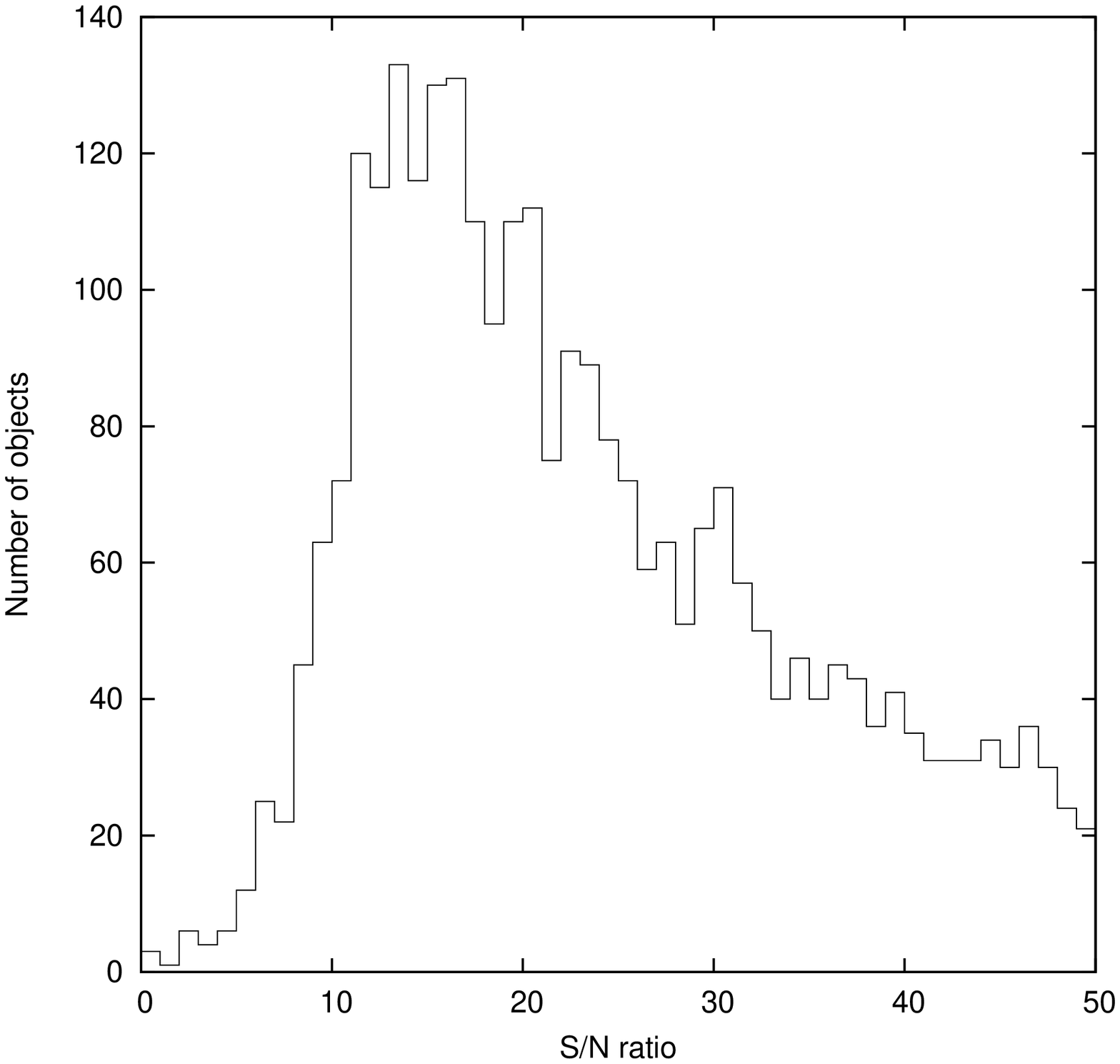}
\end{center}
\caption[]{Histogram of the signal-to-noise ratio for the Q1--Q8 KOI (first row of Table \ref{multicount} sample. Approximately 1/3 of the sample has a S/N greater than 50.}
\label{fig:snhist}
\end{figure}

The distribution of the transit S/N is shown in Figure \ref{fig:snhist}.  There is a rise in the observed number of planet candidates from a S/N of 50 to $\sim$15.  The increase is driven by the increase in the number of planet candidates towards smaller radii and the increase in \ik targets towards fainter magnitudes.  A sharp drop is observed at S/N below 15, which marks the transition where the KOI catalogue becomes significantly incomplete.   We also inspected all transit candidates with a S/N less than 15 and found convincing transit signals for all candidates with a S/N greater than 10.  \emph{Based on our observation of the S/N distribution shown in Figure \ref{fig:snhist} and inspection of the observed transit we only validate planet candidates with a S/N greater than 10.}  We expect a large number of lower S/N candidates in the range of 7.1 to 10 to still be good PCs.

\subsection{Occultation/Secondary Eclipse Search}\label{occ}

The primary signature of an EB relative to a transiting planet is the presence of eclipses of different depths due to the difference in surface brightness of the two orbiting stars.  The change in depth can be quite dramatic depending on the nature of the two stars.  As \ik photometry has high precision and a spectral bandpass extending to 850 nm, occultations or eclipses can reliably be found for companions with radii similar to Jupiter with temperatures greater than approximately 2000 K.  
For bright host stars (Kepmag $\sim$ 10) this limit can be pushed to even cooler temperatures (e.g., TReS-2b, \citet{bar12}).   As such, a secondary event can be due to a secondary eclipse from a stellar binary, or an occultation when a planet is blocked by the host star.  To distinguish between secondary eclipses and occultations, we estimated the expected equilibrium temperature (\teq) for an orbiting body heated by incident stellar flux and compared it to an estimate of the temperature (\teffp) based on the depth of the occultation.   The expectation is that a star, which is self luminous from nuclear fusion, will have a temperature \teffp\ that is much larger than \teq.  We also test whether the depth of the occultation is consistent with reflected light from a planet by computing the geometric albedo, $A_g$ in the \ik bandpass.  The secondary event is inconsistent with the planet hypothesis if $A_g$ is significantly greater than unity.

Although visual inspection reveals some obvious occultations present in the data, we performed a more thorough search to identify occultations and eclipses.  To search for secondary events the light curve was phased to the orbital period and for each phase point the mean was calculated.   Observations that occurred within 1 transit-duration were compared to mean values computed at phases within $\pm$1 transit duration.  The difference divided by the standard-deviation of observations at all phases was computed and used to identify occultations at any phase outside of transit.

To distinguish between planetary occultations and stellar eclipses, we compared the event depth with the occultation depth expected by a highly radiated exoplanet.  We estimate the equilibrium temperature by
\begin{equation}\label{eq:teq}
\teq = \teff (R_{\star}/2a)^{1/2} [f(1-A_{\rm B})]^{1/4},
\end{equation}
where \rstar\ and \teff\ are the stellar radius and temperature, $a$ is the semi-major axis, $A_{\rm B}$ is the planet's Bond albedo, and $f$ is a proxy for atmospheric thermal circulation.  To calculate the mean, we assume $A_{\rm B} = 0.1$ for highly irradiated planets \citep{row06} and $f=1$ for efficient heat distribution to the night side.  The occultation depth was used to estimate the temperature of the companion (\teffp) using our best estimate of the stellar parameters,
\begin{equation}
\teffp^4 = \teff^4 \frac{\rstar^2}{\rpl^2} \frac{F_p}{F_\star},
\end{equation}
where $F_p/F_\star$ is the ratio of the companion and stellar flux and is equal to the depth of the occultation.  We are assuming that the occultation depth observed over the \ik bandpass is a proxy for the true bolometric flux ratio.  We estimate uncertainties in \teq\ and \teffp\ by propagating our determined errors in the stellar parameters from Table \ref{spars}.  We also estimate whether the occultation could be due to reflection rather than thermal emission by estimating the geometric albedo, 
\begin{equation}
A_g=\frac{F_p}{F_{\star}}\frac{a^2}{R^2_p},
\end{equation}
In the case that \teffp\ is greater than \teq\ at the 99.7 percentile (3-sigma) and $A_g$ is greater than 1, we identify the event as a stellar eclipse and the candidate as an EB FP.  While unexpected, such a test may classify self-luminous planets (e.g., youth or external forces) as FPs.   

A number of FPs were detected through the identification of secondary eclipses (see \S\ref{fp}). The only planet in a multi-planet system with a detected occultation was Kepler-10b with $A_g < 1$ and \teffp\ $\sim$ \teq.  The lack of detected occultations in the multi-planet population is a consequence of the dearth of large-highly irradiated planets in these systems (see Figure \ref{fig:popplot}).  For the single planet population, it is likely a handful of EBs that show occultations are classified as close-in Jupiter-sized planet candidates heated to $\sim$2000 K because such planets have an occultation and transit depth similar to an eclipsing low-mass star.  The philosophy for the KOI catalogue has been to keep a candidate classified as a PC until strong evidence is presented that shows the FP nature of the candidate.  A consequence is that a handful of FPs will be misclassified as large (Jupiter-sized) candidates, which has no impact on our validation of multi-planet systems.

\subsection{Phase Linked Variations}\label{phase}

The orbital motion of a companion is imprinted in the photometric light curve due to day-night effects (thermal emission and reflectivity), ellipsoidal variations from gravity darkening due to tidal forces and Doppler boosting from orbital motion.  The latter two effects are dependent on the mass of the companion and, when present, can be used to estimate the mass of the companion (e.g., \citealt{maz13}).  

There are cases when a companion interacts with the stellar surface through magnetic fields and produces star spots that could be misinterpreted as phase linked variations due to a massive companion (e.g., Tau Bootis \citep{wal08}).  Thus our analysis would label a planet such as Tau Bootis as a FP.

To search for phase-curve variations (only relevant for short period systems), we filtered the data using the same procedure described in \S\ref{lca} except we changed the time-scale of the polynomial fitter to 5 days instead of 2 days as the filter is destructive to astrophysical signals with a similar or longer time-scale.  This means our initial search is not sensitive to phase-linked variations on these longer timescales.  The search was performed by calculating the occultation depth statistic introduced in \S\ref{occ}, which is equivalent to average filtered data with a width equal to the transit duration.  The standard deviation of the set of occultation measurements was calculated.  This value was compared to the standard deviation of the data.  This test determines whether the scatter is Gaussian on transit-duration timescales.  When the ratio was found to be greater than 2, variability in the phased light curve is detected.  In these cases we inspected the light curves and found evidence of coherent EB effects as well as variable stars with fast pulsation or rotation timescales. 

%From the multi-planet sample, Kepler-10b was the only planet to show a phase-curve.  
As was the case with the occultation search, it is expected that phase-linked variations will be rare because highly-irradiated Jupiter-sized objects are rare in the \ik multi-planet sample.  KOI-1731.02 and 1447.02 were found to show phase-curves and labeled as FPs.  From the single-planet population, phase-curves were discovered in KOI-23, 130, 143, 631, 636, 681, 699.  If ellipsoidal variations are source of the signal, then the mass of the companion was estimated which ranged from $\sim$0.1 to 0.23 \msun\ indicating that these sources are EB FPs.

\subsection{Odd-even Metric}\label{oddeven}

The occultation/eclipse search is effective when the orbital period is correctly estimated.  In cases when EB eclipses are similar in depth, it is common to have the period off by a factor of 2.  For most EBs, the depths of the alternating (odd- and even-numbered) transits differ.  We search for this odd-even effect by separately modeling the odd- and even-numbered transits where we only allowed \rprs\ to vary and the other parameters were fixed to their global solution.  We used the change in \rprs\ as a proxy for a change in transit depth.  When the change was greater than 3$\sigma$, we inspected the transit light curves to insure that the effect was real.  For the cases where we noticed spot-crossing-induced variations in the transit depths, the systems were retained as candidates. 

From the multi-planet sample, only KOI-966.01 was found to exhibit an odd-even transit effect.  Thus, the true orbital period is double the reported KOI value.  From the single-planet population 102 candidates had detected odd-even effects, although this count is incomplete as a FP is not always searched for additional effects or FP signatures in the light curve.

\subsection{Centroid Analysis}\label{centroids}

A dominant source of false positive planetary transit detections is eclipsing binaries, or giant planet transits, on background stars that are captured in the aperture of the target star (BEB).  These background signals are diluted by the target star and can have the appearance of small-planet transits.  In this subsection we describe the method we use to find KOIs with ``clean'' centroids, where the measurement is of high quality and there is no indication that the transit is not on the target star.  This ``clean'' centroid standard, described in detail below, is a more stringent centroid standard than that used for planet candidate status (see, for example, \citet{bat13}), and gives us confidence that the centroid signal is coincident on the sky with the target star.

We use centroid analysis to identify KOIs that are not clearly on the target star.  The centroid method we use is the fit of a Point Response Function (PRF) to the pixel difference image constructed by subtracting an average in-transit image from an average out-of-transit image \citep{bry13a}.  This centroid method provides an offset from the target star position for each quarter, and the final offset for the KOI is a robust average of the quarterly offsets.  We also use data quality metrics that indicate whether the data support the centroid offset measurement.  We do not validate a KOI if the centroid offset suggests that the transit is not very close to the target star, or if the data quality does not provide confidence that the transit signal is on the target star.  This centroiding method does not work for highly saturated targets.  Our treatment of saturated targets is described in \S\ref{manualInspection}.

We provide here a brief overview of the PRF-difference image centroid method.  For details see \citet{bry13a}.  PRF-based centroids are measured on both out-of-transit and difference images quarter by quarter.  When the target star is isolated, the centroid of the out-of-transit image gives the position of the target star. Assuming that the transit source is the only source of variability in the aperture, the centroid of the difference image gives the location of the transit signal source. These quarterly centroid measurements are robustly averaged to estimate the target star and centroid signal locations on the sky.  These average locations are differenced to provide the average offset of the transit signal from the target.  The robust average also provides a 1-$\sigma$ uncertainty per quarter, which is propagated through the robust average and offset calculation to provide an offset uncertainty.  An alternative method for estimating the centroid uncertainty is via a bootstrap, using a resampling with replacement of the quarterly centroid measurements.  The bootstrap-estimated uncertainty is also propagated through the robust average and offset calculation.  We choose the larger of the two offset uncertainties when performing the cuts described below.

Centroid measurements are subject to several systematic errors, caused primarily by errors in the measurement of the \ik PRF and crowding by background stars. The systematic error due to PRF error is mitigated by computing the offset as the difference between the PRF centroid of the out-of-transit image and the PRF centroid of the transit signal in the difference  image.  Because the transit source and the target star are near each other on the \ik focal plane, their PRF errors are very similar so centroid systematic errors due to PRF error approximately cancel.  The residual PRF error systematic varies from quarter to quarter and is statistically zero mean, so averaging over quarters further reduces PRF-error-driven centroid systematics.  The residual systematics have a statistical standard deviation of less than 0.1\arcsec.  To account for this systematic error, a constant $0.067$\arcsec\ is added in quadrature to the final offset uncertainty.  This added constant does not, however, eliminate all apparently significant offsets due to systematic error, so we pass any KOI with offsets less than 0.3\arcsec, even if that offset is formally statistically significant.

In a few cases there is a field star in the target's aperture that is brighter than the target.  In this case the centroid of the out-of-transit image is strongly biased by the bright star, and the centroid offsets are invalid.  We detect such cases by computing the offset of the out-of-transit image centroid from the catalog position of the target star, and declare the centroid measurement to be invalid if the offset of the out-of-transit image centroid from the target star catalog position is $\geq$  1.5\arcsec.

We classify a KOI as having ``clean centroids'' if it passes three criteria, described in more detail below: 1) it has a good centroid measurement, 2) that centroid measurement indicates small offsets from the target star, and 3) there is at least a 99\% probability that the transit signal is on the target star rather than another known star.

\paragraph{Good Centroid Measurement}\label{centroidProbability}
The quality of a centroid measurement is determined by several factors, most notably the transit S/N and systematic error.  We do not validate KOIs as planets for which difference images are not available.  There are three ways in which a KOI can \emph{fail} to have a good measurement:

\begin{itemize}

\item When the S/N is very low, the measured offset uncertainty can be too large to sufficiently localize the transit signal.  When the offset uncertainty is $\geq$ 1.5\arcsec\ we say that the KOI does not have a good measurement.
\item The measured offset of the out-of-transit centroid from the target star's catalog position is $\geq$ 1.5\arcsec, indicating that it is likely that the out-of-transit centroid measurement is strongly biased by crowding.
\item The quality of the difference image in a quarter is determined by measuring the correlation of the difference image pixels with the fit PRF.  If the correlation is less than 0.7, we consider the signal in the difference image too weak to trust the centroid value; otherwise we say that quarter has a good PRF quality.  We demand that there be at least three quarters with good PRF quality, or that with 2/3 of the observed quarters have good PRF quality, otherwise we say the KOI does not have a good measurement.

\end{itemize}

\paragraph{Small Offsets}
We demand that the measured offsets be close to the target, satisfying {\emph both} of the following criteria:
\begin{itemize}
\item The offset is statistically close, that is the offset is $< 3 \sigma$, or the offset is $< 0.3$\arcsec\ to allow for small systematic error.
\item The offset is smaller than 4\arcsec.
\end{itemize}

\paragraph{Probability $\geq 99\%$}
The systematic due to crowding is addressed via forward modeling of the observed pixels based on catalogs and the \ik PRF \citep{bry13b}.  A synthetic pixel scene is created for each quarter by placing a flux-scaled PRF at the pixel location of every known star close enough to contribute flux to the observed aperture.  In this way a synthetic pixel image modeling the average out-of-transit image is created for each quarter.  A synthetic in-transit pixel image is created for each star in the aperture by reducing the flux of that star by the transit depth that best reproduces the overall observed transit depth, accounting for dilution.  These images are analyzed for each star via difference-image PRF centroiding just like the observed pixels.  The resulting offsets provide a prediction for the transit signal offset from the target star under the hypothesis that the transit occurs on each star in the aperture.  The predicted offsets are compared with the observed offsets by inferring the underlying probability distributions.  For each star in the aperture, the normalized integral of the product of the observed and modeled distributions provides a relative probability that the transit signal is at the same location as that star, when the modeled depth on that star is less than 100\%.  An unknown background source is also included as an alternative hypothesis.  For details see \citet{bry13b}.  For this paper we assume an underlying Gaussian distribution, which is characterized by the mean and uncertainty of the offset averages. We say a KOI is not clean if its relative probability is less than 0.99.

The KOI is considered clean if the measurement quality, small centroid offset, and probability criteria are all satisfied.

\subsubsection{Manual KOI Inspection} \label{manualInspection}
Some KOIs considered in this paper do not have well-computed centroids, either because the KOIs are on saturated target stars or because the centroid did not satisfy the ``good measurement'' criterion.  Some of these KOIs were subject to manual inspection based on the criteria described in this section.  If they pass inspection, they are considered ``good".  We do not consider a ``good'' classification as strong as a ``clean'' classification, but as described in Paper II the multiple planet probability boost allows us to validate ``good'' KOIs.

\paragraph{Saturated Target Stars}
When the target star is saturated or near saturation ($Kp < 12$), centroiding methods based on the PRF are no longer valid.  In these cases the transit signal has a distinctive pattern in the difference image \citep{bry13a}.  Visual examination of the difference image in each quarter provides a qualitative indication that the transit source is in the same pixel as the target star.  Specifically, when the transit is on the target star, the transit signal appears at the end of the saturated columns, as well as in the non-saturated wings of the PRF.  We pass a saturated KOI as ``good" when the transit signal visibly appears as expected at the end of the saturated column and the transit signal wings match the non-saturated wings of the target star.  All of the saturated multis considered in this paper for which there are difference images and which are not already confirmed planets passed this test.

\paragraph{Bad Centroid Measurement}
When there are two ``clean'' KOIs in a system and additional KOIs that fail the ``good measurement'' criteria, manual inspection of the pixel data was performed to see if there is any indication that these additional KOIs are not at the target star location.  The typical situation is that the difference images were too noisy to support a high-quality centroid measurement.  In this case, when manual inspection indicates that the transit signal is on the same pixel as the target star, and that there is no significant signal in the difference image away from the target star, we consider the KOI to be ``good''.

\subsection{KOIs with Validation Issues from Imaging}

Table \ref{imaging} lists candidates that have newly-detected companions inside the photometric aperture by \citet{adams12, adams13}  and one of us (S.H.). Because these companions were not in the catalog used to compute the probability criterion described in \S\ref{centroidProbability}, we give these special attention.  Table \ref{imaging} gives the observed offset of the newly found companion from the target star and the offset of the companion from the measured transit source in units of the centroid uncertainty.  When the companion star is more than $3 \sigma$ from the measured transit source we consider that companion as ruled out as a source of the transit signal.  Because our validation criteria includes the requirement that the centroid source be no more than $3 \sigma$ from the target star, companions outside of $4 \sigma$ will not reduce the KOI's probability of being on the target star to less than 99\%.   We do not consider whether or not the companion is gravitationally bound to the primary star.

When the companion stars are within $3\sigma$ of the transit position, the transit signal is not necessarily a false positive.  However, this indicates that we did not determine which star was the source of the transits.  We do not validate such candidates unless we have strong evidence that the nearby star is a bound companion to the \ik target as described in \S 9 of Paper II. 

\subsection{Identified FPs Among the Multis}\label{fp}

All of the FP tests described above have been applied to the original sample of 1212 planet candidates identified as potential multi-planet systems.   For each FP a brief description of the types of FPs detected in multi-planet transiting systems is presented below.  The FP disposition was used for a comparison of the single planet, multi-planet, EB and FP samples.  There are three classes of FPs used to describe the nature of the transiting object: (1). Period and epoch (P/T0) collisions where multiple sources show the same orbital period and transit times.  Such events can be produced by direct PRF contamination, optical reflections or electronic interference, such as crosstalk \citep{cou13}, (2). Flux FPs, where the photometric light curve shows evidence of an EB.  (3). Active-Pixel-Offsets (APOs) FPs where centroid measurements of the photometric aperture indicate that the source of the transits is due to a source offset from the \ik target.  Categories (2) and (3) are not mutually exclusive.  Table \ref{multicount} gives a break down of the planet candidates for various cuts and number of systems with 6 candidates, 5 candidates, 4 candidates and so forth.

There were 12 P/T0 collisions detected in the following KOIs: 376.01, 489.01, 989.02, 1119.01, 1196.01, 1231.01, 1231.02, 1747.02, 1803.02, 1806.01, 1944.02, 2188.01.  A secondary eclipse was detected for most of these systems, indicating that the primary sources of P/T0 collisions are EBs.  The distribution of P/T0 collisions will not favour transit candidate targets, thus it is expected that the rate of P/T0 collisions is lower for multi-planet sample relative to the single planet sample.  KOI-489.02 was flagged a FA with a transit S/N of 6.6, thus the KOI-489 system does not count as a multi-planet system in any of our statistical counts.

%--1961.02 is low S/N, not a FP.

KOI-199.02 shows a secondary eclipse with a depth of 50 ppm and an orbital period of 8.8 days.  The depth of the secondary eclipse is inconsistent with the planet hypothesis. Derived values from the occultation are $A_g\sim22$, \teff $\sim$ 4500 K and \teq=1200 K. 

KOI-376.01 is a P/T0 collision and shows strong quarterly depth variations due to quarterly variation of contamination.  The second candidate, KOI-376.02, has a period of 1.4 days and shows a strong secondary eclipse with an observed depth of 260 ppm.  Since the signals observed are likely heavily diluted, the true eclipse depths are likely much deeper.

KOI-379.01 was found to have an additional star within the photometric aperture with a separation of 1\arcsec.  Centroid analysis points towards the fainter star as the source of the transits, thus this candidate has been flagged as an APO.  Centroids analysis of KOI-379.02 is inconclusive to determine which star is the source of the transits and is kept as a PC.

KOI-414.01 shows a secondary eclipse with a depth of 400 ppm.  The location of the secondary eclipse indicates that the orbit is non-circular.  KOI-414.02 shows a clear centroid offset and was flagged as an APO.  

KOI-2671.01 was marked as a FP as a secondary eclipse was detected.  The occultation shows the orbit to be eccentric (35 hours offset).  KOI-2671.02 has a centroid offset and was labeled as an APO.

KOI-989.01 and KOI-989.02 are the same event with the periods being integer multiples of one another.  These two candidates were also flagged as a P/T0 collision.  The confusion of these two candidates arose because of strong variations in the quarter transit depths from quarterly dependence of dilution.  KOI-989.03 remains as a PC.

KOI-549.01, 549.02, 1196.01, 1231.01, 1378.02 and 2007.01 are flagged as APOs from centroid analysis.   KOI-1119.01 is a P/T0 collision and 1119.02 shows strong centroid offsets. KOI-1342.01 shows a small offset of 0.9\arcsec\ with a significance of 4.1$\sigma$. It is therefore considered to be an APO. KOI-2159.02 shows centroid offsets and a secondary eclipse.  KOI-1731.02 shows an occultation and phase-linked variations and was labeled as a FP because its transit depth appears to be heavily diluted.

KOI-966.01 shows an odd-even transit effect, but KOI-966.02 was labeled as a FA due the low S/N of the transit event.  Thus, the KOI-966 system does not count as a multi for our statistics.

KOI-1447.01 showed a `V' shaped transit event with a depth greater that 15\%.  While transit-depth is not an indication of the FP nature of the candidate, KOI-1447.02 shows large amplitude phase-linked variations.  Thus, KOI-1447.02 is a clear FP which removes the KOI-1447 system as a multi-planet system, and due to the large transit-depth we classified 1447.01 as a FP.

%KOI-1915.02 was labeled as a FP in the Q1--Q8 catalogue paper, but I don't think this is true.

From the single-planet population of 2482 PCs, 976 were classified as FPs resulting in a FP rate of $\sim$40\%, which, as expected, is in stark contrast to the multi-planet FP rate.   In total we found 26 FPs  (including P/T0 collisions) in the multi-planet sample of 1167 PCs remaining after the removal of FAs and single planets.  A few of the classified FPs had an associated FA, such as KOI-1231.02, and are included in the FP totals for the single planet population.  The 26 FPs include cases of two FPs associated with the same target.  Candidates that were flagged as having {\it not clean} centroids in Table \ref{mplanetfit} are not FPs and remain as unvalidated PCs.  There is no strong evidence to suggest that any one target with {\it not clean} centroids is a blend, however, the probability of blends existing within the population is large enough that we cannot validate this sample at the 99 percentile.  The 26 FPs are found around 20 systems, with 6 double FP systems, 12 cases of FP associated with a single PC and 2 cases where 2 PCs and 1 FP where associated with the same target.  

The results of the multi-planet disposition are summarized in Table 4 of Paper II, which gives a comparison and breakdown of the expected FP rate.  The agreement between the observations and predictions is very good which leads to our conclusion that a vast majority of transiting candidates found in multi-planet systems are genuine planets.  After removal of FPs there are 1129 remaining multi-planet candidates.  The next step is to explore this large population of candidates and set additional criteria to reduce the chances that undetected blends still exist.  There will be a population of FPs from blends that exist in the \ik transit sample, but cannot be detected via our methods, for example, a blend from a BEB where the separation from the target source is too small to be detected by centroid motion.  

\subsection{Validation of Multi-Planet Candidates}\label{vps}

To reduce the number of potential FPs in our validated list of planets in multi-planet systems, we eliminate regions of phase space where we have reduced confidence.  For example, we have reduced confidence in the validity of a PC if centroids cannot localize the position of transit to eliminate the chance of a background blend at the 99 percentile.  The first requirement is that a candidate has a S/N $>$ 10 as established in \S\ref{snr}.  This insures that the multi-planet sample is free of FAs and removes an additional 21 candidates after the removal of FPs and P/T0 collisions.  

\begin{figure}
\begin{center}
\includegraphics[scale=0.5]{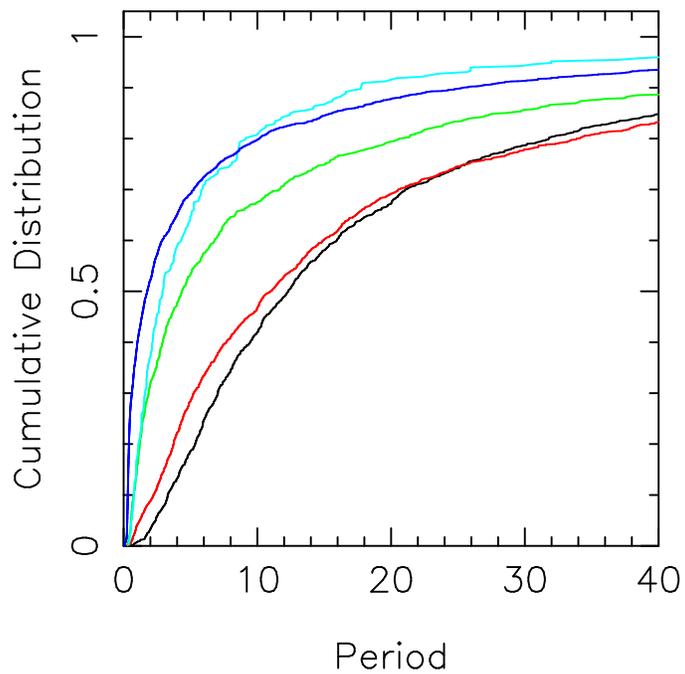}
\end{center}
\caption[]{Cumulative distribution of orbital periods for multi-planets (black), single KOIs (red), FP KOIs (green), P/T0 collisions (cyan) and EB systems (blue).  All periods used in normalization.  Other time scales are shown in Figure 2 of Paper II.}
\label{fig:perdisp}
\end{figure}

Using the analysis for \S3, \S4 and this section, we are able to use the criteria set out in Paper II to select a population of multi-planet transiting systems that have a FP rate substantially less than 1\% (additional details below).  Table \ref{multicount} lists the number of planet candidates after various cuts and tests are applied.  The various cuts are: {\bf FA}, where either a transit candidate has insufficient S/N ($<$ 7.1) or was labeled as a non-transit event such as stellar variability or was observed to have less than 3 transits. {\bf Col} indicates a P/T0 collision.  These sources are non-unique by nature so they are classified separately.  {\bf FP}  indicates when a FP is identified that is a not a P/T0 collision.  A FP can either be an EB masquerading as a planet candidate or a diluted signal where the source of the transits has been localized off the \ik target.  {\bf SN} The transit models were used to determine the S/N of the phase folded transit for each candidate.  We adopted a threshold of S/N $>$ 10 to consider a transit event.  {\bf P} marks period cuts.  We require the orbital period to be greater than 1.6 days due to the increased rate of FP found with shorter orbital periods (See Figure \ref{fig:perdisp} in \S\ref{fp} and Appendix A of Paper II).  {\bf b} marks when cuts are made based on the measured impact parameter.  When a transit is `V' shaped there is a larger chance that a FP has been identified.  This does not mean that `V' shaped events are FPs, only that we have less confidence in declaring such objects as planets.  The fraction of `V' shaped signatures that are produced by EBs as opposed to transiting planets is far larger than that for `U' shaped profiles.  Our criteria for `V' shaped transits is that $b$ + $b_{\sigma}$ + \rprs\ $>$ 1.00.   {\bf centr} indicates the centroid test as outlined in \S\ref{centroids}.  A target that does not pass our centroid test is not a statement that the object is a FP/APO, but rather that we had insufficient information to localize the source of the transits on the \ik target. The column {\bf \# of multi-pl} indicates the total number of multi-planet systems that pass the indicated test.  The column {\bf \# of new multi-pl} indicated the total number of multi-planet systems that pass the indicated test and have been previously verified (already assigned a Kepler-ID).   

Figure \ref{fig:perdisp} shows the cumulative distribution of orbital periods for candidate multi-planet systems (black), candidate single planet systems (red), FPs KOIs (green), P/T0 collisions (cyan) and EBs from the Kepler Eclipsing Binary catalogue\footnote{http://keplerebs.villanova.edu/}.   FA are not considered.  The distribution of EBs shows a large population of short period events due to the inclusion of contact binaries.  The sample of FPs from the KOI catalogue also shows a relatively strong population of FPs at short orbital periods, which is different from the EB population, but there is still a much larger fraction of FPs at shorter orbital periods compared to the PC populations.   There are two reasons for this difference: (1) KOIs are selected based on a visual inspection of the photometric transit-event, with a requirement that the event has an appearance of a planetary transit.  This process heavily reduces the number of contact binaries in the KOI catalogue, as a distinct transit that shows a clear ingress and egress are not present.  (2) The \ik transit search algorithm does not conduct searches for events with periods less than 0.5 d, so unless a strong harmonic of the orbital period is detected at a longer period, many short period events will also be missed.   Similarly, the distribution of P/T0 collisions, which is dominated by EBs, shows a larger fraction of events with short orbital periods relative to the PC distributions.  As articulated in Appendix A of Paper II, the expected abundance of unidentified FPs in multis to planets in multis is far larger at small orbital periods than at large ones.  Thus we do not validate PCs with orbital periods less than 1.6 days. We also excluded systems with an orbital period less than 4 days and a S/N $<$ 15, due to concerns of an increased FP rate and the lack of well constrained transit model parameters due to the low S/N of the transit.  Eliminating candidates with a $P$ $< $1.6 days reduced the sample of 1107 PCs in multis that passed our FP and S/N tests to 1084 PCs as shown by row 6 of Table \ref{multicount}.  

Dispositioning of PCs relies on transit models to characterize the orbiting companion.  A transit can be `V'-shaped in appearance when: the transit duration is comparable to the photometric cadence or we have a grazing transit ($b$ + \rprs\ $\sim$ 1).  The transit model incorporates the cadence time to convolve the synthetic lightcurve to match observations and allows a quantitive assessment as to whether a grazing transit is observed.   A grazing transit also results in increased uncertainty in transit model parameters, which makes assessment of the transit event difficult.  From the single transit population, after the removal of FAs, 16\% have grazing or close to grazing transits ($b$ + $b_{\sigma}$ + \rprs\ $>$ 1.00), which drops to 8\% after the removal of FPs and P/T0 collisions, which is larger than one would expect based on an isotropic distribution of orbital planes.   Almost half of the grazing transit single PCs have a radius greater than 15 \rearth, which indicate that a large fraction of this population are likely FPs.  However, planetary radius is not a criterion to label a candidate as FP because an upper limit on the radius of a planet is not well established, and measurement uncertainty on `V'-shaped transits precludes making a definitive statement regarding the absolute radius of the transiting object.  From the multi-planet population, after the removal of FAs,  3.8\% are measured to be grazing or near grazing, which drops to 3.1\% after the removal of FPs and T/P0 collisions.  Only 1 (5.9\%) of the grazing multi-planet candidates has a large inferred planetary radius, KOI-1477.01, which is associated with an EB (KOI-1477.02).  It is very likely that KOI-1477.01 is also an EB FP.  The lower rate of grazing transits in the multi-planet population leads us to conclude that a large fraction of the grazing transits in the single-planet population are FPs.  This means that there is a higher probability of a FP being found when a grazing transit is present, thus we do not validate multi-planet candidates that have $b$ + $b_{\sigma}$ + \rprs\ $>$ 1.00.  These KOIs are still good PCs.  Application of FP, P/T0, S/N and cuts based on impact parameter reduces the set of multi-planet candidates for validation to 1054.

Cuts based on S/N, period and transit-shape use the transit models and comparison with the single planet population to identify regions of model parameter space with reduced confidence in the validity of a planet candidate; when the rate of FPs is observed to be larger relative to the multi-planet population.  Cuts based on methodology presented in \S\ref{centroids} where PRF models were used to identify which multi-planet candidates have at least a 99\% probability that the transit signal is on the target star rather than another known or unknown star.  This statement means that there is a low probability that a blended background transit event is present.  After application of our centroid criteria the number of multi-planet candidates that is still considered for validation is 851.
%This criterion is very conservative as it does not consider the additional requirement that not only is a background star present, but that target has a transiting companion that can produce the observed transit shape.    

As previously mentioned, planet radius is not a criterion for classification of a transiting candidate as a FP.   A problem thus arises for large Jupiter-sized planet candidates as there is a degeneracy in radius for planets, brown dwarfs and low mass stars.  Additional dynamical tests were applied to the multi-planet candidates: using Hill's criterion to test for stability of neighboring pairs of planet candidates, and when \rpl + 2$\sigma_{R_{\rm p}}$ $>$ 9 \rearth, dynamical fits  to transit times observed in Q1 -- Q14 \ik long cadence data were conducted to determine whether the giant candidates can have masses exceeding 13 \mjup, assuming all of the candidates orbit the \ik target.  Details of both tests can be found in Appendix C of Paper II.  All candidates that passed the above tests, apart from the special case of KOI-284 (Kepler-132, addressed in \S 9.1 of Paper II), passed the stability tests.  Table \ref{mplanetfit} categorizes candidates as having passed, failed or being to small to have been tested for mass limits large enough to be stars.  The last row of Table \ref{multicount} gives the number of planetary candidates after all tests have been applied and gives us a total of 851 planets that we validate.  From this sample, 60 have been previously validated via other methods, thus {\bf we are able to introduce 768 newly validated planets, which roughly doubles the current number of confirmed and validated planets.}  Planets discovered and confirmed by the \ik mission currently account for more that half of the known and validated extrasolar planets.

\section{Population of Validated Planets and Discussion}\label{pop}

\begin{figure}
\begin{center}
\includegraphics[scale=0.75]{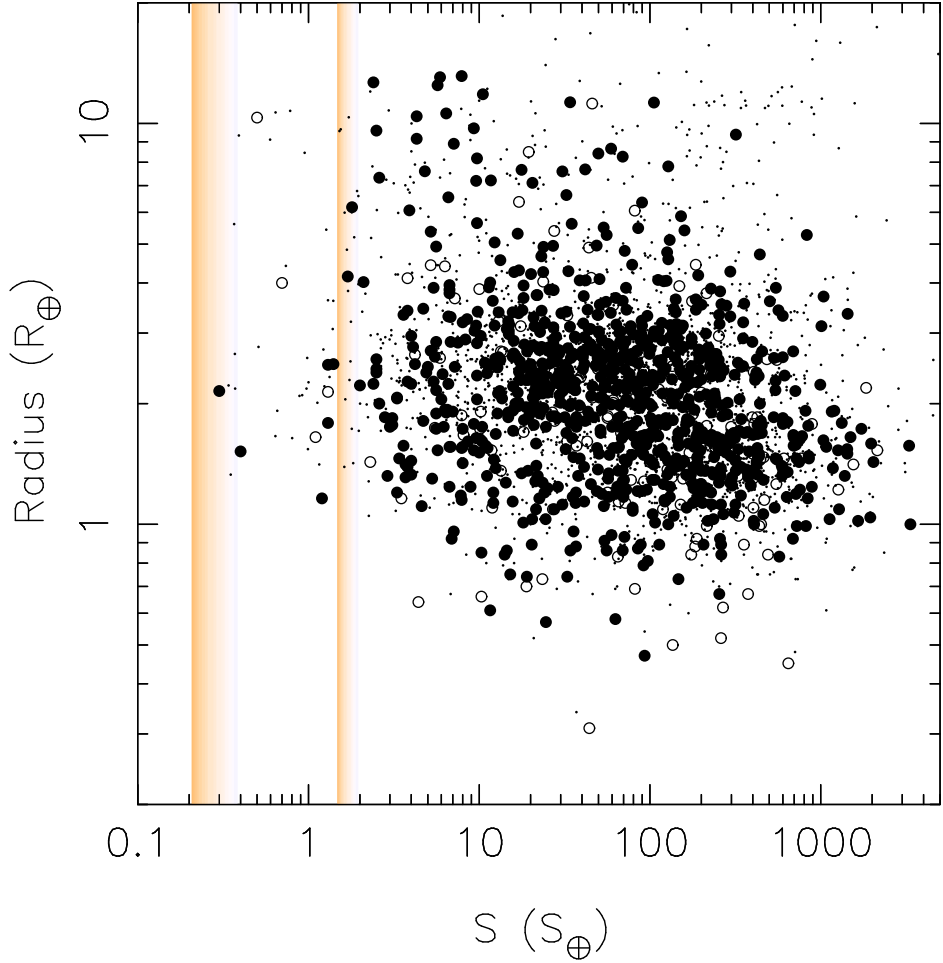}
\end{center}
\caption[]{Multi-planet population showing incident flux $S$ (relative to the flux received by the Earth), versus planetary radius (\rearth).  Planets validated in this paper are marked with filled circles, unvalidated planets in multi systems are marked with open circles and planet candidates found in single systems are plotted with dots.  All planets and planet candidates shown have $P > 1.6$ days, $b+ b_{\sigma}$\rprs $< 1.00$ and have not been identified as a FA, FP or P/T0 collision.  The two coloured bands display estimates of the inner and outer habitable zone based on the Recent Venus and Early Mars models from \citet{kop13}.  The colours display the range each boundary as a function of \teff\ from 3000 K (red) to 7000 K (light-blue).}
\label{fig:popplot}
\end{figure}

After the application of all the tests listed above, we validate 851 extrasolar planets associated with 340 planetary systems.  These systems are expected to have a FP rate that is significantly less than 1\% due to the reasons listed in \S\ref{vps} and the theoretical framework laid out in Papers I and II.  From this sample, 768 candidates in 306 systems have not been previously validated, but are now extrasolar planets validated above a confidence level of 99\%.  Thus, we introduce Kepler-100 through 405.  From this population there are 106 new planets that have a radius less than 1.25 \rearth\ compared to 16 that have been previously validated.  There are 6 planets with incident solar flux, $S$, less than 1.5 times that of the Earth including four planets: KOI-518.03, 1422.04, 1430.03 and 1596.02 (Kepler-174 d, Kepler-296 f, Kepler-298 d, Kepler-309 c respectively) that are new validations that we discuss below.  Figure \ref{fig:popplot} plots incident flux versus radius and displays our new validations as filled circles. Multi-planet candidates that pass all tests except our centroid criterion are plotted as open circles and single planet candidates after the exclusion of FPs are plotted as small dots.   The falloff in the number of planets below 1 \rearth\ is driven by incompleteness due to insufficient S/N.  The falloff in the number of  planets with $S < 1.5$ is due to decreasing transiting probability and incompleteness to longer period events ($>$150 d)  as our sample is based on Q1-Q8 photometry ($\sim$ 2 years).  As noted by \citet{lat11}, there is a lack of hot-jupiters in multi-planet systems verified in Figure \ref{fig:popplot} as a deficit of planets with $S>200$ and \rpl\ $> 7$ \rearth\ relative to the single planet population.

The FP tests presented in \S\ref{fpi} are not sensitive to most hierarchical blends.  A hierarchical blend is a bound stellar binary with a transiting planet orbiting one of the stellar components.  It is not known if widely separated binaries host planetary systems with orbital planes aligned with the stellar orbital plane.  If the alignment distribution is isotropic, then hierarchal, transiting triples in the multis may be rare, however, if alignment is common, say because of star-planet formation processes that favor aligned systems, then the rate of hierarchal, transiting triples could be much larger.  For an isotropic distribution it was shown in \S5 of Paper II that 4 or 5 systems are likely to have a planet candidate around each stellar component and it is very unlikely that there is more than 1 hierarchical triple multi (3 bound stars, one hosting a transiting planet and the other 2 in a eclipsing configuration).  If the orbital planes of planets around both components of a stellar binary are aligned then we might expect to find a greater number of blends.  In Appendix A, we develop a synthetic population model to test whether hierarchical blends could contribute a large fraction of the observed multi-planet population by comparing the measured value of \rhostar\ from our transit models.  We find that the multi-planet population is not dominated by hierarchical blends, but the strongest constraints come from KECK HIRES observations to search for spectroscopic blends (\S 8.1 of Paper II).  Together, it appears the rate of hierarchical blends is low.  It is important to note that even if any of our validated planets are found to be orbiting a fainter and bound star, they are still planets; however, the stellar parameters listed in Table \ref{spars} will need to be revised.

The single and multi-planet populations also appear to have different fractions of planets at longer orbital periods. The relative cumulative distribution of the multiple planet systems overtakes the single planet populations at a period of $\sim$25 days.  There are 1027 and 897 single and multi-planet candidates with periods between 5 and 150 days.   If we separate these samples into short periods, 5-25 day, and long periods, 25-150 days, we find 689 and 627 short period planets and 338 and 270 long period planets for the single and multi-planet populations respectively.  Thus, 55\% of the multi-planet sample are found in the short period bin compared to 45\% for the single planet sample.   Explaining this difference seems counterintuitive, as alignment of orbital planes in multi-planet systems would make it more likely to find longer period planets relative to the single planet population under the assumption that long period planets are equally common in multi and single planet systems.  However, there are strong biases in the detection process that generates the raw KOI list. In particular the candidates are found via different numerical and inspection methods.  For example, there are a number of long period single-transit candidates that were found through identification of a single transit event and then the candidate was continuously monitored for additional transits. 

\subsection{New Planetary Systems with Planets In or Near the Habitable Zone}

We discuss herein, validated multiple planet systems that contain a planet that is in the nominal habitable zone of their star.  The location of the habitable zone depends on stellar luminosity (and the orbital period range also depends on stellar mass), so we introduce only those planets whose host stars have been characterized spectroscopically in this section.  %Additionally, if evidence for additional stars is present in the target's spectrum or in high-resolution imaging, we ** may exclude the planet unless transit durations imply that it is highly unlikely that the planet orbits the target or it is in HZ for either star**

\begin{figure}
\begin{center}
\includegraphics[scale=0.50]{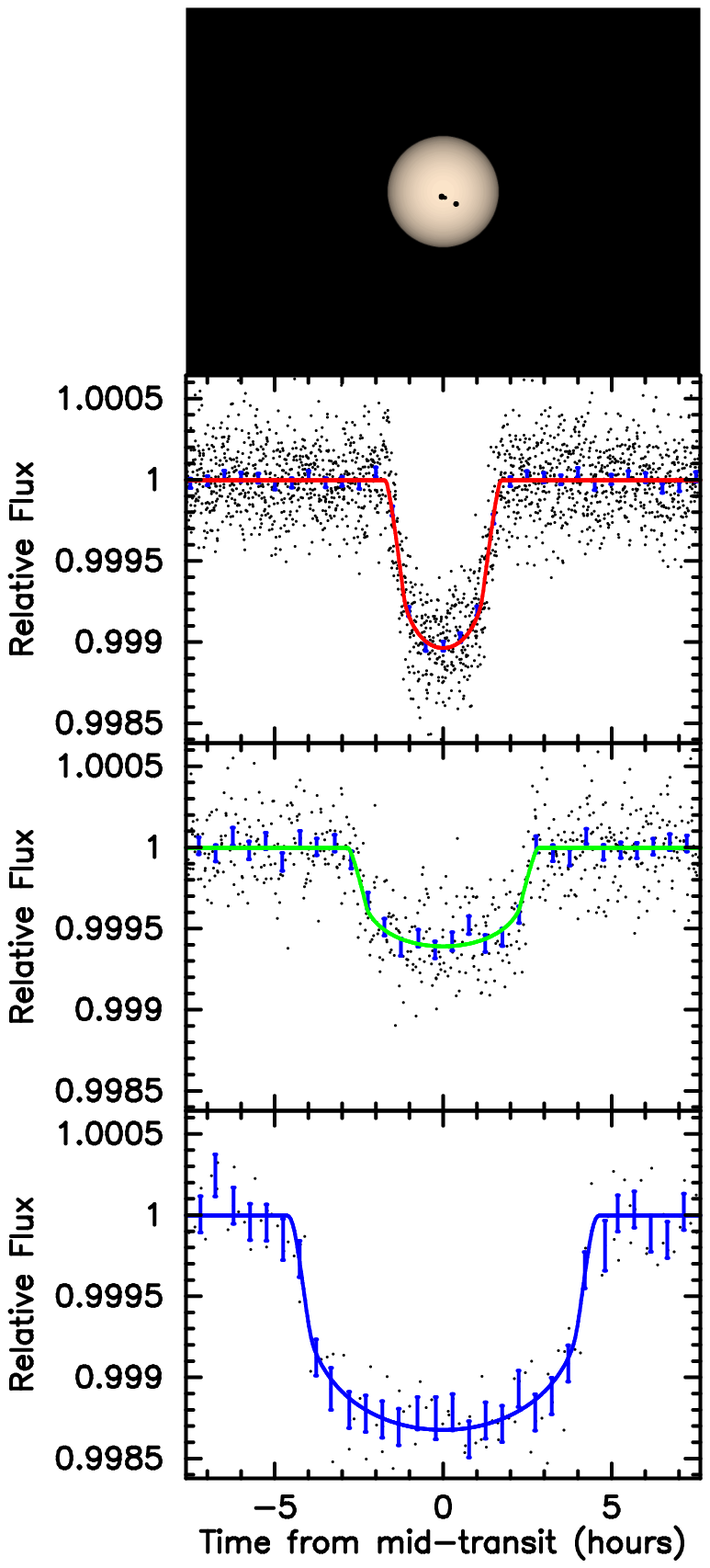}
\includegraphics[scale=0.50]{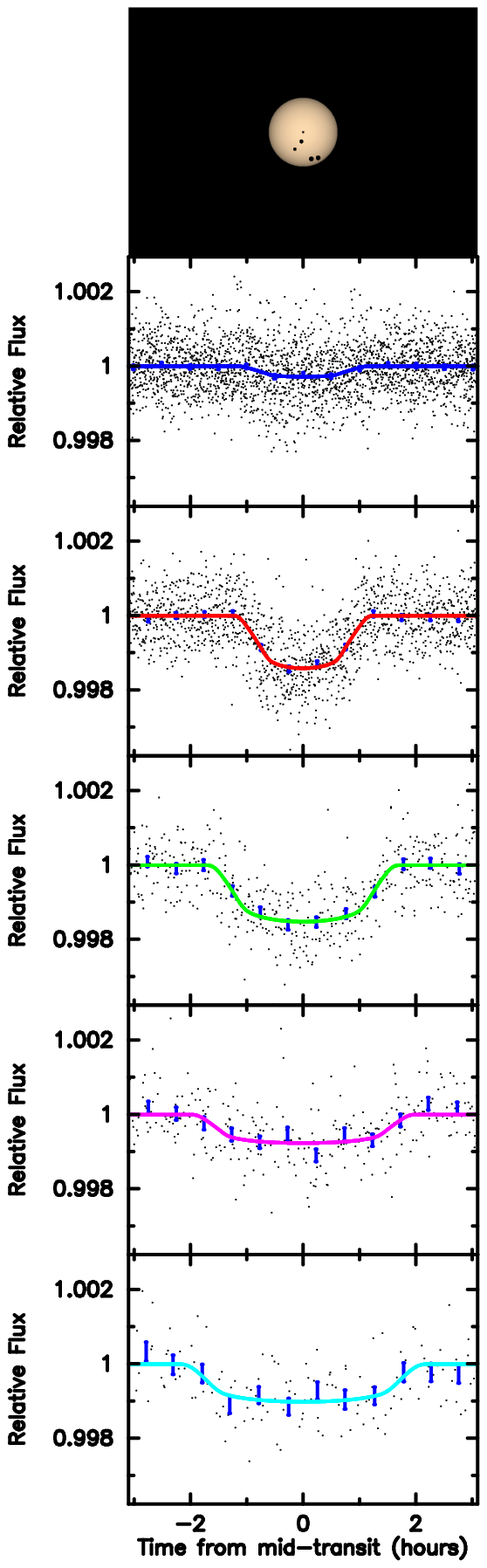}
\includegraphics[scale=0.50]{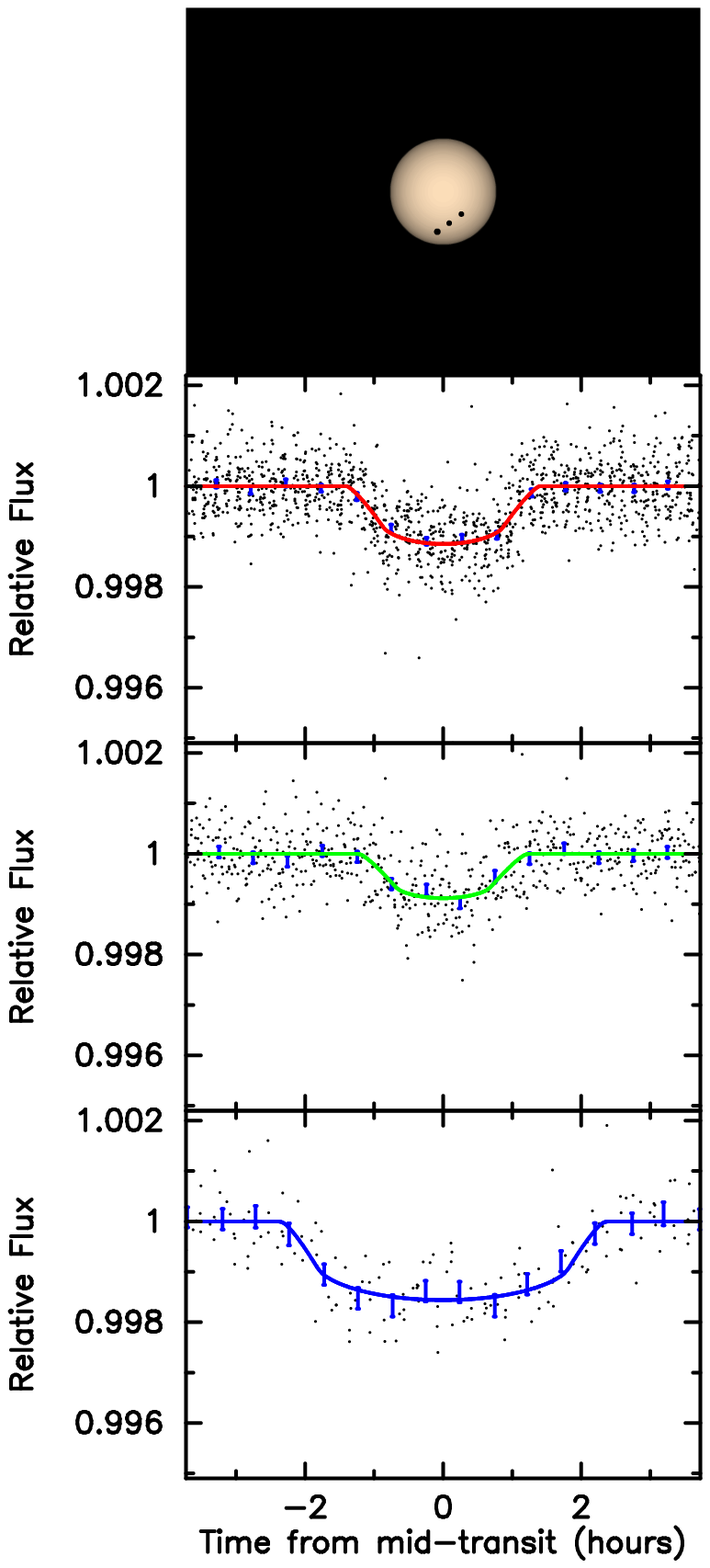}
\includegraphics[scale=0.50]{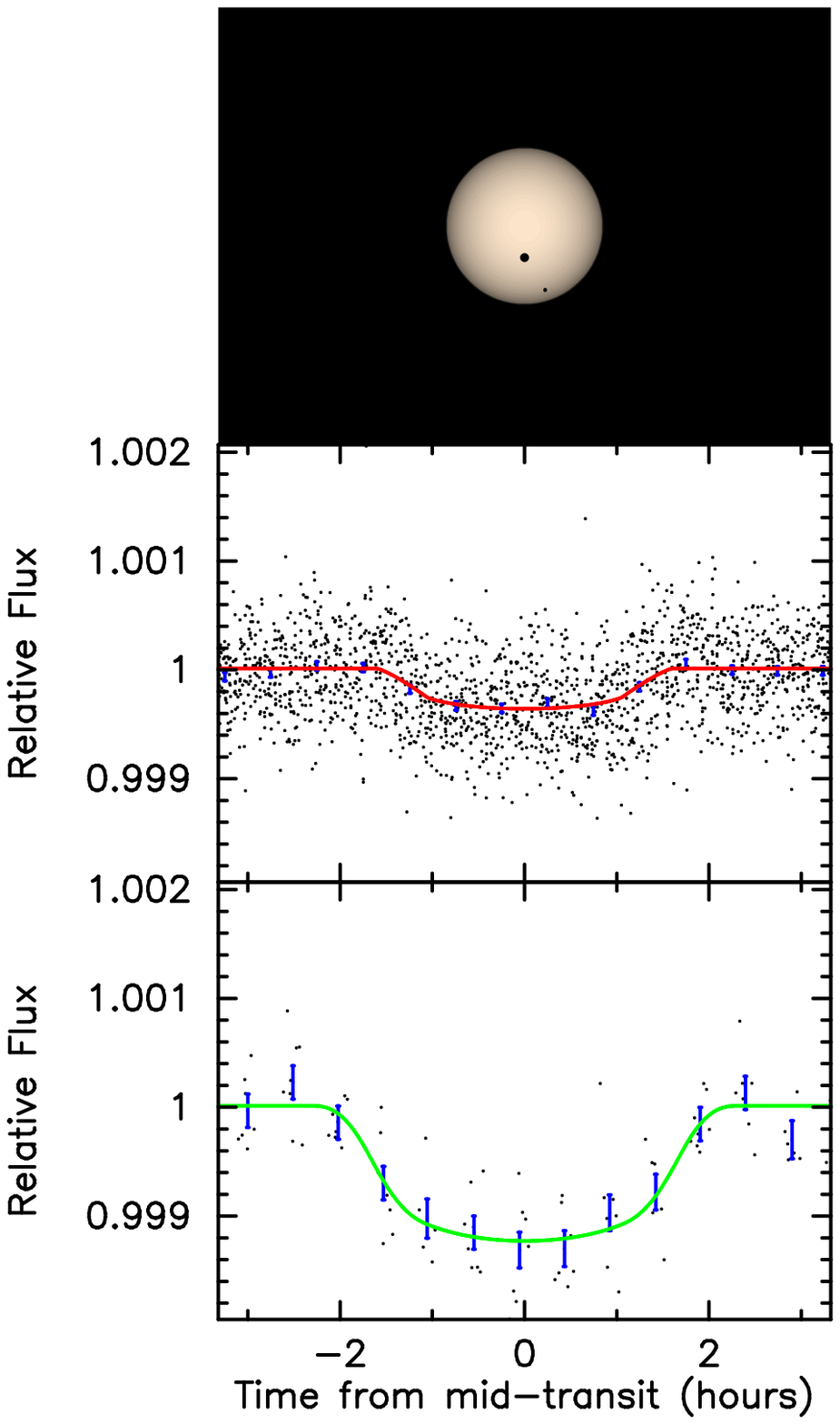}
\end{center}
\caption[]{Transits of all of the validated planets in systems with a newly validated planet with  $S <$ 1.5. The beauty shots in the upper panels display the sizes of the stars and planet candidates to a uniform scale. The color of the stars and the impact parameters of the planetary transits reflect estimates of stellar and transit characteristics given in Tables 2 and \ref{mplanetfit}. Verified planets are shown in black while other candidates are green. Planets and candidates are displayed with distance below the middle of the star corresponding to the transit impact parameter. The lower panels show the detrended Kepler flux from the host star phased at the period of each transit signal and zoomed to a region around mid-transit, shown in order of increasing orbital period. Black dots represent individual Kepler long cadence observations. The blue bars are the data binned 30 minutes in phase with 1$\sigma$ uncertainties. The colored curves show the model transit fits, with colors corresponding to the last two digits of KOI designators as follows: red = .01, green = .02, blue = .03, cyan = .04. In each panel, the best-fit model for the other planet candidates was removed before plotting. All panels for a given system have an identical vertical scale, to show the relative depths, and identical horizontal scale, to show the relative durations, but scales differ between systems. The successive panels show KOIs 518, 1422, 1430 and 1596.}
\label{fig:tplots}
\end{figure}

As we do not have information regarding either the albedo or atmospheric characteristics of these planets (nor of any moons that they might have), we can only make reasonable estimates of the flux of stellar radiation that they intercept, i.e., the amount of insolation that they receive.  We therefore quote results in terms of the average solar flux intercepted by Earth, $S$, which is generally referred to as the solar constant.  For the purposes of our tabulation, we list objects that intercept flux less than 1.5~$S$ (comparable to the flux Venus received 1 billion years ago, see \citep{kop13} and references therein). For comparison, Venus receives 1.91~$S$.  We don't specify an outer boundary to the HZ because few of the \ik planets that we have validated have significantly smaller insolation than does Earth, but note that Mars receives an average of 0.43~$S$. Orbital eccentricity, $e$, which generally is unknown, affects the flux of stellar radiation that a planet receives, but the change that it induces in the annual average insolation  is roughly  quadratic in $e$, and as few of the planets in \ikt's multis have large eccentricities, the magnitude of the change in mean insolation resulting from planetary eccentricity is likely to be small.  The spectrum of stellar radiation received by a planet also affects atmospheric and surface temperatures \citep{kas93}, but these variations are small compared to uncertainties in estimated insolation and in atmospheric properties.  Nonetheless, we note the effective temperatures of the stellar hosts to aid in investigations by other researchers.  Only six of the planets that we validate intercept less than 150\% of the radiation flux encountered by Earth.  Two of these orbit KOI-701 (Kepler-62) and have been analyzed in detail by \cite{Borucki:2013}.  The transits for the four new planets that receive less than 1.5~$S$, KOI-518.03 (Kepler-174 d), 1422.04 (Kepler-296 f) (see Paper II for more detailed discussion), 1430.03 (Kepler-298 d) and 1596.02 (Kepler-309 c), are shown in Figure \ref{fig:tplots}.   The planets have nominal radii of 2.19, 1.79, 2.50 and 2.51 \rearth.  

\subsection{Conclusions}

Our work provides a substantial increase in the number of verified exoplanetary systems and demonstrates the ability of the \ik mission to probe the statistics of exoplanetary systems with a sample that is relativity clean of FPs.  Both transit models and centroid models are used to characterize the photometric data and various tests were used to identify FPs.  The rate of FPs was found to be low relative to the single transiting planet population, in quantitative agreement with theory (Papers I and II). This result demonstrates that 851 planet candidates in multi-planet systems are valid planets at greater than the 99\% level.  In Appendix A, the multi-planet population was used to investigate the rate of hierarchical blends in multi-planet systems; while no limits on the occurrence rate can be currently set, it was found that the eccentricity distribution of transiting multi-planet systems found by \ik is significantly different from the planetary distribution found by RV surveys.  The list of validated planets presented is reliable, but the sample suffers from both incompleteness and strong biases.  Many of the candidates that were not validated in this study are still excellent planetary candidates.  
%As additional planet candidates are discovered with \ik photometry, additional multi-planet systems will be discovered and subject to validation.

\acknowledgments

Funding for this Discovery mission is provided by NASA's Science Mission Directorate.  We are indebted to the entire {\it Kepler} Team for all the hard work and dedication have made such discoveries possible. JFR is partially supported by a Kepler Participating Scientist grant (NNX12AD21G).

\appendix

\section{Rate of False Multis from Transit Durations}\label{tdur}

The FP tests presented in \S\ref{fpi} are not sensitive to all types of hierarchical blends.  We consider a {\it hierarchical blend} to be a gravitationally bound stellar binary with a transiting planet orbiting one of the stellar components.  In this section we lay out the framework for estimating the number of blends in the \ik multi-planet sample.  Our aim was to set limits on the rate of hierarchical blends, which could be a large source of error in the Kepler exoplanet database.  We present two types of analysis: 1. comparison of measured values of \rhoc\ for each planet pair within a multi-planet candidate system.  2. comparison of \rhoc\ to \rhostar\ derived from stellar theory.

Our transit models provide a measurement of \rhoc, which is the measurement of the mean stellar density, \rhostar\, if the transiting planet travels on a circular orbit.  The value of \rhoc\ is strongly correlated to the transit duration, which in turn depends on the planet's semi-major axis and the impact parameter for a circular orbit.  Each transiting planet in a hypothetical planetary system with planets in perfect circular orbits will produce the same measurement of \rhoc.   Variations in the measured value of \rhoc\ can be produced in three ways:  (1). Eccentric orbits change the transit duration depending on the star-planet separation at the time of the transit in accordance with Kepler's Second Law. Comparison of a population of transits provides some insight on the eccentricity distribution because \rhoc\  and \rhostar\ will differ.  (2).  Unresolved, gravitationally-bound stars that host transiting planets that have diluted transits as both stars are observed within the same photometric aperture.   The unseen companion is fainter and, generally, smaller relative to the target star.   Thus, hierarchical triples produce systematically larger values of \rhoc.   When the planet system is bound to the fainter star, the measurement of \rhoc\ can disagree with the estimate of \rhostar\ from stellar classification and modeling.  If transiting planets are found around both components of the stellar binary, then \rhoc\ will not agree planet to planet.  (3). Measurement error introduces scatter that produces \rhoc\ values that are equally too small or large.  The population of \ik\ multi-planet systems was used to place limits on the rate of hierarchal blend occurrences as described below.

The difference between values of \rhoc\ for each KOI was computed for each planet candidate in the system.  For example, if a system has 3 planets (P1, P2 and P3), then we would compare the difference in \rhoc\ based on P1-P2, P1-P3 and P2-P3 relative to the sum of \rhoc\ for each pair,
\begin{equation}
\drho = \left| \frac{\rho_{c,i}-\rho_{c,j}}{\rho_{c,i}+\rho_{c,j}} \right| .
\end{equation}
Since the distribution is symmetric, we used the absolute value.   The binned distribution of \drho\ measurements is shown in black in Figure \ref{fig:drhodisp}.   The sample included 1158 planet candidates after the removal of FAs, FPs and P/T0 collisions.  The observed distribution is broad but peaked towards zero.   The next step was to construct a synthetic population to reproduce the observed distribution that accounted for eccentricity, binarity and measurement error.  

%The distribution of $<$\rhostar$>$ versus \drho\  for each planet pair in Figure \ref{fig:drho}, where $<$\rhostar$>$ is the average of $\rho_{\star i}$ and $\rho_{\star j}$.   We only plot multi-planets KOIs with at least two planets that have a S/N is greater than 7.1 and not labeled as a FP or FA.  The colours represent the total number of planets in the system.

\subsection{Constructing Synthetic Populations}

To construct a synthetic population, \rhostar\ for the primary star was adopted from Table \ref{spars} and matching orbital periods for the orbiting planets from Table \ref{mplanetfit}.  Thus, a synthetic transiting planet is generated for each planet in our transiting multi-planet sample.  A co-eval binary companion is generated for each star and later we decide whether the primary or secondary component is hosting the planet.  The companion is not used to estimate the binary fraction, but to determine the change in \rhoc\ and estimate the number of hierarchical blends.   A hierarchical blend occurs when transiting planets are observed around both components of a stellar binary.  Masses for a bound companion ($M_2$) were chosen to be greater than 0.1 \msun\ and less than the primary ($M_1$) and have a mass-ratio ($q$) distribution,
\begin{equation}
{\rm N}(q) \propto q^{n},
\end{equation} 
where n=-1 would produce a 1/$q$ distribution matching the distribution observed in radial velocity surveys \citep{tri90}.   The transit depth from Table \ref{mplanetfit} together with an estimate of the luminosity of the primary ($L_1$) and secondary ($L_2$) stellar components was used to check that the undiluted transit-depth around the fainter secondary star would not exceed 50\%.  This sets a lower limit on the luminosity of the secondary, 
\begin{equation}
L_{2,min} = 2 L_1 \rm T_{dep},
\end{equation}
where T$_{dep}$ is the transit depth from Table \ref{mplanetfit}.  If $L_{2}$ $<$ $L_{2,min}$, then we choose another mass for the stellar companion and repeated until a suitable choice is found for this check.  To determine which stellar component the planet would be orbiting in our model, a fitted parameter,  {\it binfrac}, was used to represent the fraction of planets that are orbiting the fainter, bound companion.  For each transiting planet in the system we drew a uniform random number.  If that number was less than {\it binfrac} then we adopted \rhostar\ of the bound companion.  

A system has now been constructed that consists of two bound stars, each having a probability of having a transiting planet.  To account for eccentricity in the synthetic population, a two-parameter model of the eccentricity distribution based on the beta function, as described in \citet{kip13}, was adopted,
\begin{equation}\label{beta}
P_{\beta} = \frac{1}{\beta(x,y)} e^{x-1} (1-e)^{y-1},
\end{equation}
where $e$ was the eccentricity of the planet and $x$ and $y$ were fitted parameters.   Other distributions, such as a Rayleigh distribution, could also be used to produce similar results.   The distribution $P_{\beta}$ was used to draw a value of $e$ for each transiting planet.   With the orbital period and \rhostar\  selected, \adrs\ was calculated using Equation (\ref{rhostar}). The argument of periapsis ($\omega$) was then randomly selected from a uniform distribution and used to determine the star-planet separation, $d$/\rstar, during transit.   The transit probability was then calculated,
\begin{equation}
T_{prob} = \frac{\rstar}{d},
\end{equation}
and a uniform random number from 0 to 1 was drawn.  If the random number was greater than $T_{prob}$ then the choice of $\omega$ was rejected and a new value was drawn and the exercise repeated.  This process insures that transits occurring near periastron are preferred.  The estimate of d/\rstar\ was substituted in Equation (\ref{rhostar}) to estimate \rhoc,
\begin{equation}\label{rhostarc}
\left( \frac{d}{\rstar} \right)^3 \simeq \frac{\rhoc G P^2}{3 \pi}.
\end{equation} 
Measurement error was incorporated by choosing a model solution from the MCMC analysis for the corresponding KOI and comparing \rhoc\ to the median value of \rhoc\ from all the chains.  The difference was added to the synthetic value of \rhoc.  To investigate the dependance of the synthetic model on reliability of estimating uncertainties in transit parameters, we used a nuisance parameter, {\it errfrac}, to scale the errors on \rhostar\ as measured by the transit when assuming a circular orbit. 

\subsection{Results}

\begin{figure}
\begin{center}
\includegraphics[scale=0.75]{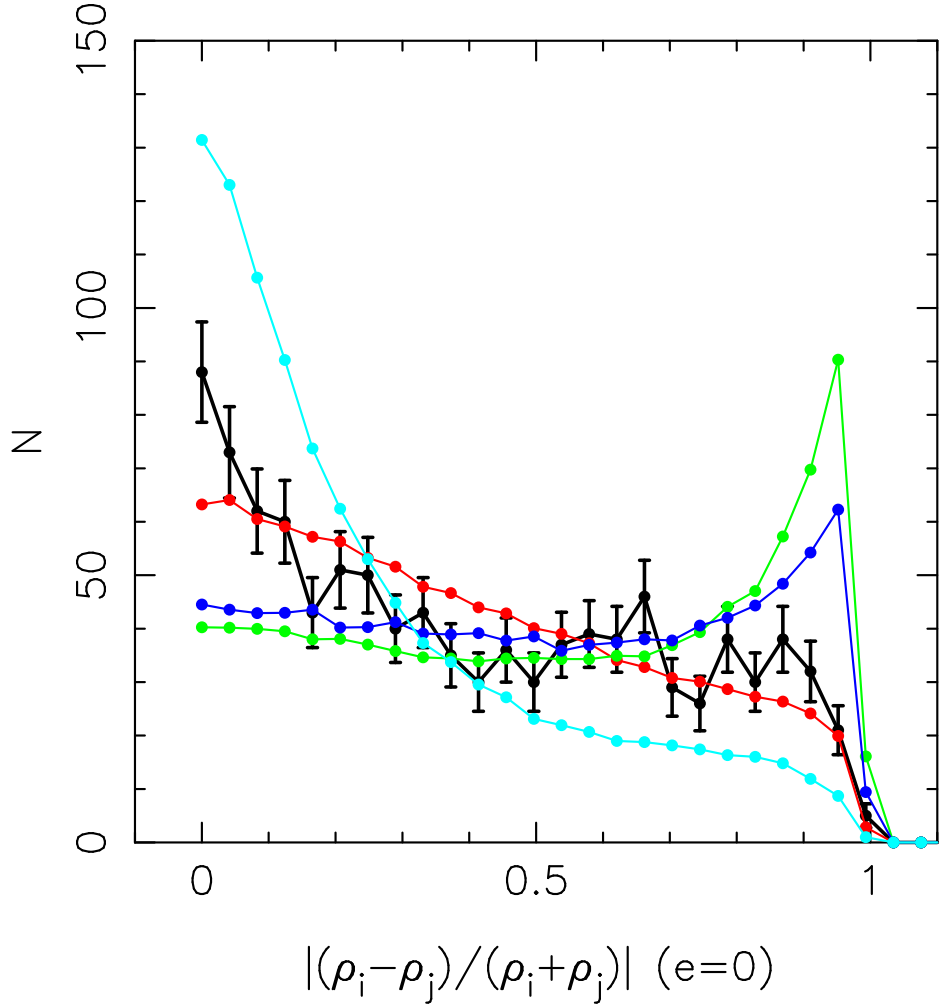}
\end{center}
\caption[]{The distribution of differences in \rhostar\ as measured by two planets orbiting the same star.  The observed \ik\ multi-planet population is shown in black with 1 $\sigma$ Poisson uncertainties.   Other lines show synthetic population models.  All models incorporate the measurement uncertainty of \rhostar\ from Table \ref{mplanetfit}.  The cyan line shows a synthetic model population with circular orbits and no hierarchical blends.  The red line shows a synthetic population with an eccentricity distribution based on RV detected planets.   The green line shows a synthetic population with eccentricity and a hierarchical blend rate of 0.5.  The stellar companion masses are drawn from a 1/$q$ distribution.  The blue line shows a synthetic population similar to the green line except the stellar companion masses are drawn from a uniform distribution. }
\label{fig:drhodisp}
\end{figure}

In Figure \ref{fig:drhodisp} we plot the binned distribution of \drho\ based on various synthetic populations for comparison to the observations.   The cyan line shows a population of planets with only circular orbits and no hierarchical blends; only incorporating measurement error.  This model does not match the observations shown in black.  The red line was produced by incorporating measurement error and an eccentric distribution of planets from radial velocity planets \citep{wri13} with best fit parameters from \citet{kip13}.  This model produced a better fit, but not an ideal fit to the observations.  The green line shows a population produced using measurement error, circular orbits and a hierarchical blend rate of 0.5 and N($q$) $\propto$ $q^{-1}$.  In this scenario half of the planets are transiting primarily low mass stellar companions.  The blue line shows a population with circular orbits and a hierarchical blend rate of 0.5 and a uniform distribution of $q$.  For both cases with hierarchical blends we see an overabundance of mismatches in \rhoc\ between planet pairs, with the strength of the mismatches modulated by the distribution of $q$.  

\begin{figure}
\begin{center}
\includegraphics[scale=1.00]{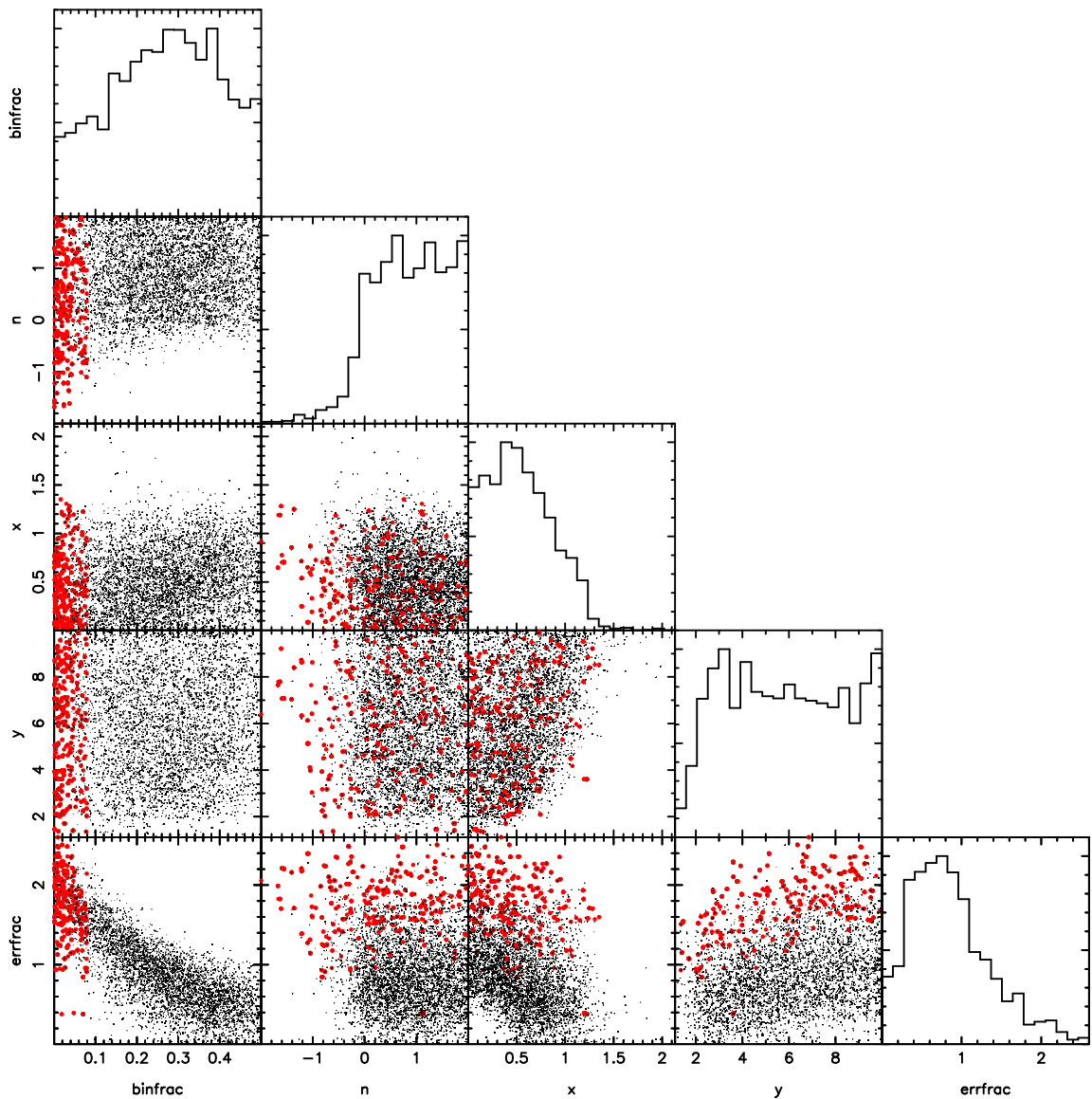}
\end{center}
\caption[]{Plots of MCMC distributions for parameters describing a synthetic population of planet systems to measure the rate of hierarchical blends.  The diagonal panels show histograms for each parameter, other panels show scatter plots of two parameters against one another.}
\label{fig:tdurplots}
\end{figure}

To measure posterior distributions of the parameters to describe the planet population we use a MCMC routine that uses methods similar to the description found in \S\ref{mcmc}.   In Figure \ref{fig:tdurplots} we show distributions for  {\it binfrac}, n, a, b and {\it errfrac} based on 48 000 chains.   Parameters where restricted to  {\it binfrac}=\{0,0.5\}, n=\{-2:2\}, a,b=\{0,10\} and {\it errfrac} $>$ 0.   It is immediately clear that posterior distributions for each fitted parameter are quite broad, but we can draw a few conclusions. 

An extensive search for blended companions based on KECK high-resolution spectra was described in \S8.1 of Paper II.  The sample included 270 multi-planet candidate systems and would be sensitive to blends due to companions that are 2-3\% as bright as the companion and show a RV difference of $\sim$ 10 km/s.  From this sample, only 1 clear blend was found and in that case (KOI-2311.02) the S/N of the transit was found to very low, so we do not even consider that system to be a multi-planet candidate.  However, based on the one potential blend detection an estimate of the blend rate is $0.004 \pm 0.004$ for companion stars that are 2-3\% as bright as the primary.  Beyond $\sim$ 5 AU, the RV component of the stellar binary will be too small to allow reliable detection of stellar blends, which would account for less than half of companions found in solar-type stars \citep{rag10}.   We double the potential blend rate and take the 3-sigma upper limit to get a rough estimate of the number of hierarchical blends that could exist.  The synthetic population has 1158 planets in 460 systems, so observations suggest that no more than 14 blended systems could exist and either be missed by the spectroscopic survey or have separations large enough that a companion would not be detected.  The inclusion of high resolution imaging observations would allow additional constraints on the number of companions detected at large separations.  In Figure \ref{fig:tdurplots}, all simulations that have less than 14 blended systems (e.g., at least one planet around each stellar component) have been marked in red.

The rate of hierarchical blends, {\it binfrac}, was found to be dependent on how well uncertainties are determined for \rhoc.   The transit model used a fixed set of limb-darkening parameters, thus we expect our uncertainties to be both underestimated and potentially systematically biased.   We do not expect our uncertainties to be under-estimated.  There is weak (2 $\sigma$) evidence that there are zero hierarchical blends in our sample based on the measurement uncertainties in Table \ref{mplanetfit}.   Limits on the blend rate from spectroscopic analysis also suggests that the uncertainties on \rhoc\ have been underestimated. 

\begin{figure}
\begin{center}
\includegraphics[scale=0.75]{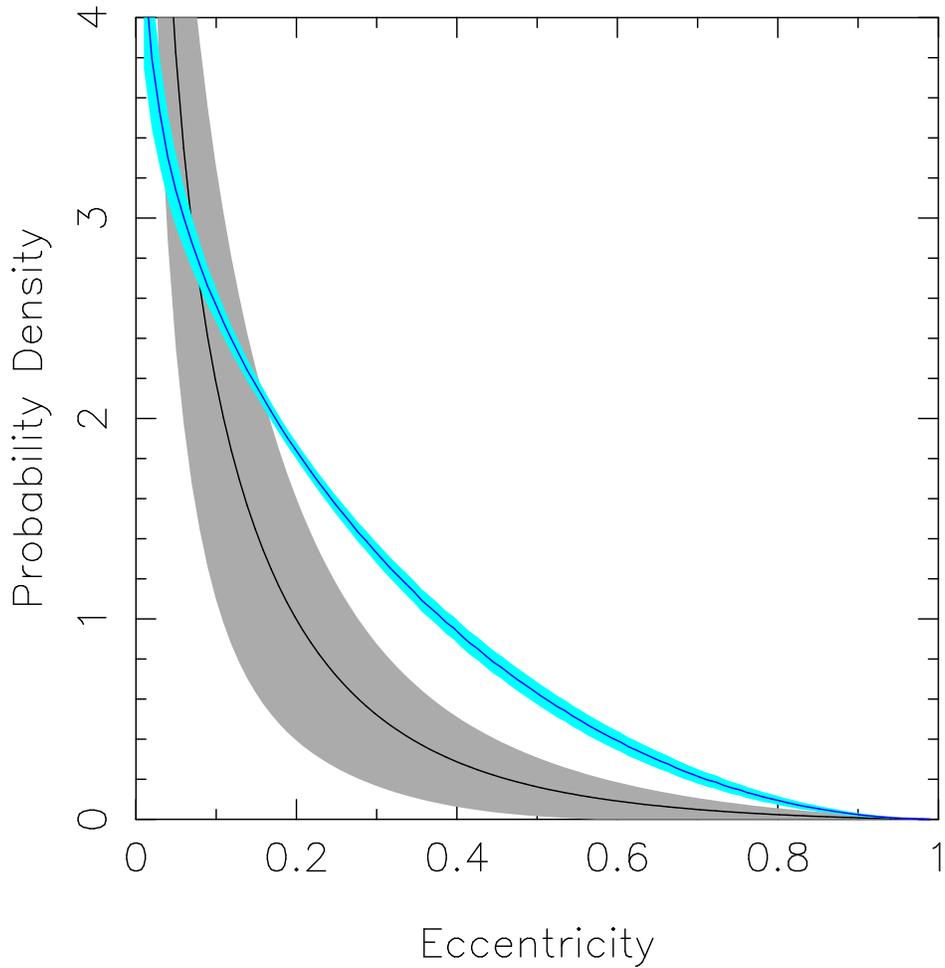}
\end{center}
\caption[]{Comparison of the eccentricity distribution of multi-planet systems (black), based on matching synthetic population models to the transiting multi-planet sample, to the eccentricity distribution of RV planets (blue) \citep{wri13,kip13}.  The distributions are different at high significance.  The relative fraction of low eccentricity planets was found to be large for the transiting multi-planet population.}
\label{fig:eccndisp}
\end{figure}

The eccentricity distribution as parameterized by Equation (\ref{beta}) is relatively independent of {\it binfrac}, {\it errfrac} and $n$.  Figure \ref{fig:eccndisp} shows the range of allowed eccentricity distributions in black based on the synthetic populations from the MCMC analysis with 1 $\sigma$ uncertainties.  The blue line shows the eccentricity distribution based on RV surveys \citep{wri13} based on an analysis by \citet{kip13}.  A chi-square test of the two distributions gives $\chi^2 = 46.2$ for 10 samples, which indicates that the two distributions are different with high confidence.  The multi-planet eccentricity distribution is more sharply peaked towards zero, thus high-eccentric planets in multi-planet systems are relativity rare compared to the RV sample.  It would be interesting to further break down the RV sample to compare the eccentricity distributions of single and multi-planet samples, but it outside the scope of the initial analysis presented here.

The difference in \rhoc\ when hierarchical blends are present will be strongly dependent on probability distribution of the underlying mass function.  As shown in Figure \ref{fig:drhodisp}, the distribution of \drho\ will contain an increased number of large mismatches as $n$ decreases.  In the case of equal or nearly equal-mass binaries the difference in \rhoc\ will be indistinguishable from measurement error.  The synthetic population model shows that as {\it errfrac} is reduced, $n$ pushes towards large values that produce a large number of equal mass binaries.  For {\it errfrac}=1, there is no strong measurement of the mass fraction distribution.

\begin{figure}
\begin{center}
\includegraphics[scale=0.75]{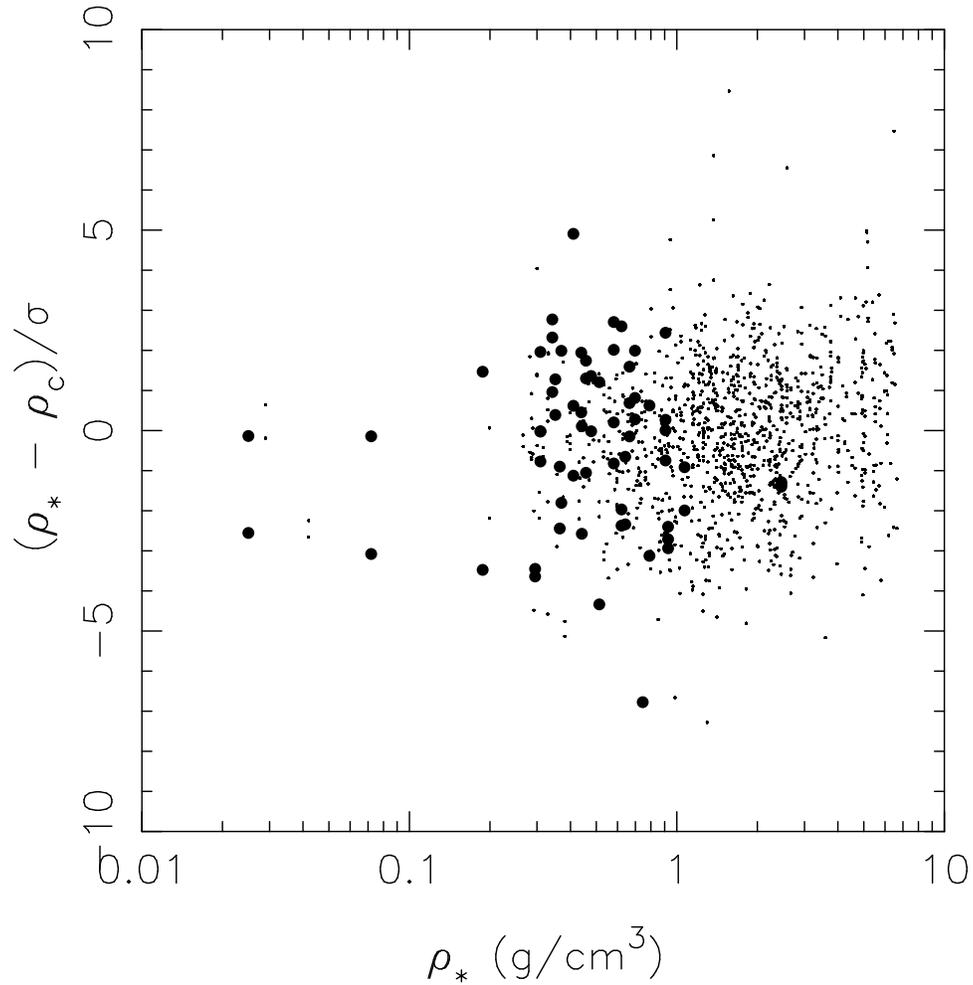}
\end{center}
\caption[]{The difference between \rhostar\ and \rhoc\ is plotted vs \rhostar\ to compare the mean stellar density derived from stellar modeling (\rhostar) to the geometrical estimate of the mean stellar density when a circular orbit is assumed (\rhoc).   Values of \rhostar\ derived from asteroseismology are shown with filled circles, values of \rhostar\ derived from fitting stellar evolution models to spectroscopic classification (see \S\ref{stellarpars}) are shown with dots.  The population of dots shifts towards positive values of \rhostar\ -- \rhoc\ as \rhostar\ increases due to a poor match of our adopted stellar evolution models at low stellar mass.}
\label{fig:rhodiff}
\end{figure}

The value of \rhoc\ can also be directly compared to \rhostar\ (i.e. \citealt{tin11}) from Table \ref{spars}.   Eccentric planets will be seen when the transit occurs close to pericenter, which decreases the transit duration relative to a circular orbit.  This pushes \rhoc\ towards larger values (a denser star).  As stated above, hierarchical blends will also produce a bias towards larger values of \rhoc\ as \ik planet host stars are typically close to the main-sequence.  Measurement error should not introduce any bias.  A comparison of \rhostar\ and \rhoc\ in a similar manor to the comparison of \rhoc\ for planet pairs could be carried out, but a strict requirement is that \rhostar\ is a good estimate of the true mean-stellar density, which is not true.  In particular, the use of Yale-Yonsei evolution models are known to produce radii too large and hence, densities that are systematically too small for low mass stars \citep{pla12}.   However, when \rhostar\ is based on asteroseismology \citep{hub13}, such biases are likely better controlled.  Figure \ref{fig:rhodiff} shows the difference in \rhostar\ and \rhoc\ scaled by the uncertainty versus \rhostar.  The asteroseismology sample is limited to solar-like and evolved stars as the amplitudes of p-mode oscillations scale proportionally to stellar luminosity.  The bias towards smaller values of \rhostar\ for low mass stars with \rhostar\ based on Yonsei-Yale models can be seen for stars with \rhostar\ $ > 3$ \gcmc.  From this small sample, there is evidence of a bias of \rhoc\ being larger that \rhostar; however, the sample is too small to draw any inferences on the underlying eccentricity and hierarchical blend population. 

Currently, the sample of multi-planet systems and well characterized stars is too small to draw strong conclusions about the number of hierarchical blends.  The strongest constraints currently come from observations that attempt to directly detect a nearby companion that may be gravitationally bound.  When additional multi-planet candidates and observations become available, model fits can be repeated to determine the rate of hierarchical blends and in turn, establish whether orbital planes in binary stars systems are aligned.

\section{List of symbols and abbreviations} 

\begin{itemize}
\item \mjup, \rjup\ -- mass and radius in Jupiter units
\item S/N -- signal-to-noise ratio.
\item KOI -- Kepler Object of Interest 
\item FA -- False-alarms are candidates with a S/N $<$ 7.1
\item FP -- Astrophysical False-positive
\item P/T0 -- Period/T0 collision False-positive produced by the instrument
\item PC -- planetary candidate
\item EB -- eclipsing binary
\item BEB - background eclipsing binary
\item $S$ - ratio of incident flux relative to the Earth.
\item \rhostar\ -- mean stellar density
\item \rhoc\ --- transiting derived mean stellar density for circular orbits
\end{itemize}

%Comparing \rhostar\ based on stellar evolution models or transit-durations from our e=0 models we find that \rhostar\ is overestimated for cooler, higher density stars which agrees with (paper from nexsci).  We have not investigated other stellar model (e.g. Dartmouth).

%improvements include injecting transits into the photometry to recover - with proper S/N using information from asteroseismology about host star.
 
%To estimate the rate of planets found around only a fainter companion...  we only use the asteroseismology sample as it gives the most precise values of \rhostar\ and excluded fainter cool stars where stellar theory and measurements can disagree (e.g. P)  

%\subsubsection{KOIs to fix for modelling and various Notes}
%KOI1196 is an EB, not a multi.  Period .01 should equal period .02 
%KOI1231 - depths change a lot for planet .01 - check centroids - probably third light
%KOI2671 - one quarter has shallow transits for .01
%KOI1447 - .01 is very V-shaped, can't seem to get a proper circular model.  Trying an eccentric model instead.  .02 has a definite phase curve, so this is likely just an EB blend.   Have a converged e=0 fit in n1.dat.  Use the linpack modeller for n2.dat and see if we can get an eccentric model to fit this thing.  Nope, just can't get a fit.

\newpage

%
%\begin{deluxetable}{cccc}
%\tabletypesize{\scriptsize}
%\tablecaption{Targets and KOI discovery Rate}
%\tablewidth{0pt}
%\tablehead{
%\colhead{\# of Quarters}   & \colhead{\# of Targets} & \colhead{\# of Multi-KOIs} & \colhead{\%}
%}
%\startdata
%1 &  5648  &   0 & 0.00 \\
%2 &  4738  &   1 & 0.02 \\
%3 & 5366  &   1 & 0.02 \\
%4 & 9869  &  11 & 0.11 \\
%5 & 11404  &  23 & 0.20 \\
%6 & 11027 &   22 & 0.20 \\
%7 & 55851 &   156 & 0.28 \\
%8 & 86848 &    252 & 0.29 \\
% \\
% Total & 190751 & 466 & 0.24 \\
%\enddata
%\tablecomments{\# of Quarters observed from Q1 -- Q8.  Only included multi's that have been examined as part of this study.  FPs are included but FA are excluded. The last columns shows the percentage of KIDs that where found to be multi's in this study (again, FAs have been excluded).}
%\label{targetcount}
%\end{deluxetable}

\begin{deluxetable}{cccccccccl}
\tabletypesize{\scriptsize}
\tablecaption{Multiplanet Counts}
\tablewidth{0pt}
\tablehead{
\colhead{S1} & \colhead{M1}  & \colhead{2} & \colhead{3} & \colhead{4} & \colhead{5} & \colhead{6} & \colhead{\# Multi-pl} & \colhead{\# new Muti-pl} & \colhead{Cuts}
}\startdata
2527 &   0 & 313 & 107 &  45 &  13 &   3 & 1210 & 1092 &                  No Cuts \\
2482 &  17 & 300 & 104 &  43 &  13 &   3 & 1167 & 1049 &                       FA \\
2161 &  21 & 295 & 102 &  43 &  13 &   3 & 1151 & 1033 &                   FA,Col \\
1861 &  27 & 293 & 100 &  42 &  10 &   3 & 1122 & 1011 &                 FA,Col,P \\
1496 &  24 & 284 & 102 &  43 &  13 &   3 & 1129 & 1011 &               FA,Col,FPs \\
1377 &  29 & 285 & 100 &  42 &  10 &   3 & 1106 &  995 &             FA,Col,FPs,P \\
1492 &  32 & 279 & 100 &  43 &  13 &   2 & 1107 &  989 &            FA,Col,FPs,SN \\
1366 &  41 & 273 &  98 &  40 &  13 &   2 & 1077 &  964 &          FA,Col,FPs,SN,b \\
1374 &  37 & 281 &  96 &  43 &  10 &   2 & 1084 &  973 &          FA,Col,FPs,SN,P \\
1257 &  46 & 272 &  96 &  40 &  10 &   2 & 1054 &  947 &        FA,Col,FPs,SN,P,b \\
--   &  66 & 236 &  81 &  32 &  12 &   1 &  909 &  821 &                  FA,centr \\
--   &  66 & 236 &  81 &  32 &  12 &   1 &  909 &  821 &              FA,Col,centr \\
--   &  67 & 235 &  81 &  32 &  12 &   1 &  907 &  819 &           FA,Col,FP,centr \\
--   &  67 & 235 &  81 &  33 &  11 &   1 &  906 &  818 &        FA,Col,FP,centr,SN \\
--   &  72 & 233 &  79 &  34 &   9 &   1 &  890 &  803 &      FA,Col,FP,centr,SN,P \\
--   &  79 & 227 &  78 &  32 &   9 &   1 &  867 &  784 &    FA,Col,FP,centr,SN,P,b \\
--   &  83 & 221 &  78 &  31 &   9 &   1 &  851 &  768 &    Dynamical and SP Cuts \\
\enddata
\tablecomments{Number of Mulitplanet Systems after various cuts.  See \S 5.9 for detailed definition of the cuts.}
\label{multicount}
\end{deluxetable}

\begin{landscape}
\input{spars.tex}
\end{landscape}

\begin{landscape}
\input{multifit.tex}
\end{landscape}

\begin{deluxetable}{ccccc}
\tabletypesize{\scriptsize}
\tablecaption{Close Neighbours Discovered with High-resolution Imaging}
\tablewidth{0pt}
\tablehead{
\colhead{KOI}   & \colhead{Companion Offset (arcsec)} & \colhead{Transit Position $\sigma$ (arcsec)} & \colhead{Companion Offset from Transit Position in $\sigma$} & \colhead{Disposition}
}
\startdata
112.01 & 0.1        & 0.1840 & 1.0 & fail  \\
112.02 & 0.1        & 0.2900 & 0.4  & fail \\
270.01    & 0.05       & saturated           & $<$3 & fail \\
270.02    & 0.05       & saturated           & $<$3 & fail \\
285.01    & 1.44       & saturated           & $<$3 & fail \\
285.02    & 1.44       & saturated           & $<$3 & fail \\
279.01    & 0.922      & saturated           & $<$3 & fail \\
279.02    & 0.922      & saturated           & $<$3 & fail \\
\enddata
\tablecomments{KOIs with otherwise clean centroids that have close companions revealed by high-resolution imaging.
For non-saturated targets, the companion offset in $\sigma$ is companion's offset divided by the uncertainty in the centroid measurement for that KOI.
When the companion is within $3 \sigma$ of the measured transit location we do not validate the target.  Two distances in $\sigma$
are given for KOI 307 because there is a 180 degree ambiguity in the position angle of the companion.}
\label{imaging}
\end{deluxetable}

\input{ttvtable.tex}

\newpage

%\begin{figure}
%\begin{center}
%\includegraphics[scale=0.5]{mescomp.eps}
%\end{center}
%\caption[]{Comparison of 3 hour CDPP for all \ik targets (black) versus those targets showing KOIs (red). As expected, KOIs are preferentially found around targets with less noise, however, the differences between the two distributions are small.}
%\label{fig:cdppdist}
%\end{figure}

%, purple = .05 and yellow = .06

%\begin{figure}
%\begin{center}
%\includegraphics[scale=0.9]{tplots6.eps}
%\end{center}
%\caption[]{Example Figure for 6 planet Systems: KOI157, 351 and 505}
%\label{fig:157}
%\end{figure}

%\begin{figure}
%\begin{center}
%\includegraphics[scale=0.75]{tdurcomp20130906p1.eps}
%\end{center}
%\caption[]{The left panel shows the \ik\ multi-planet population when comparing measurements of \rhostar\ between paris of planets within the same system.  The plot shows mean value of \rhostar\ for the planet pair versus the scaled difference between the pair of \rhostar\ .  The total number of planets with a system is represented by the colour, with red designating 2 planets, 3 planet systems as green, 4 planet systems as blue, 5 planet systems as cyan and 6 planet systems as magenta.}
%\label{fig:drho}
%\end{figure}

\end{document}

%% file: spars.tex
\begin{deluxetable}{ccccccccccccccc}
\tabletypesize{\scriptsize}
\tablecaption{Stellar Parameters}
\tablewidth{0pt}
\tablehead{
\colhead{KOI}  & \colhead{Kepler-ID} & \colhead{KID} & \colhead{\teff} & \colhead{\teff$_{\sigma}$} & \colhead{\logg} & \colhead{\logg$_{\sigma}$} & \colhead{\feh} & \colhead{\feh$_{\sigma}$} & \colhead{\rstar} & \colhead{\rstar$_{\sigma}$} & \colhead{\rhostar} & \colhead{\rhostar$_{\sigma}$}  &\colhead{Flag} &\colhead{blend} \\ 
\colhead{}  & \colhead{} & \colhead{} & \colhead{K} & \colhead{K} & \colhead{cgs} & \colhead{cgs} & \colhead{} & \colhead{} & \colhead{\rsun} & \colhead{\rsun} & \colhead{\gcmc} & \colhead{\gcmc} & \colhead{} & \colhead{}
}
\startdata
41 & Kepler-100 & 6521045 & 5825 & 75 & 4.125 & 0.045 & 0.02 & 0.10 & 1.490 & 0.035 & 0.457 & 0.013 & 5 & 3 \\
46 & Kepler-101 & 10905239 & 5570 & 134 & 4.065 & 0.240 & 0.30 & 0.10 & 1.666 & 0.415 & 0.351 & 0.300 & 3 & 4 \\
70 & Kepler-20 & 6850504 & 5443 & 74 & 4.398 & 0.100 & 0.00 & 0.07 & 0.986 & 0.095 & 1.304 & 0.400 & 4 & 3 \\
72 & Kepler-10 & 11904151 & 5627 & 44 & 4.342 & 0.046 & -0.15 & 0.04 & 1.056 & 0.021 & 1.068 & 0.008 & 5 & 3 \\
82 & Kepler-102 & 10187017 & 4908 & 74 & 4.640 & 0.100 & 0.08 & 0.07 & 0.716 & 0.032 & 3.132 & 0.304 & 4 & 3 \\
85 & Kepler-65 & 5866724 & 6169 & 50 & 4.236 & 0.035 & 0.09 & 0.08 & 1.424 & 0.024 & 0.621 & 0.011 & 5 & 3 \\
89 & & 8056665 & 6688 & 342 & 4.059 & 0.150 & -0.21 & 0.10 & 1.773 & 0.357 & 0.329 & 0.186 & 3 & 3 \\
94 & Kepler-89 & 6462863 & 6184 & 83 & 4.196 & 0.068 & 0.11 & 0.07 & 1.486 & 0.139 & 0.543 & 0.127 & 4 & 3 \\
\enddata
\tablecomments{Flag: -1 - solar parameters assumed ; 0 - original KIC ; 1 - Revised KIC ; 2 - SPC ; 3 - Specmatch ; 4 - SME ; 5 - Asteroseismology \\
blend flag: 0 - nearby star detected that may produce blend, 1 - no measurement, 2 - has speckle, 3 - has spectral matching, 4 - has both}
\label{spars}
\end{deluxetable}

%% file: ttvtable.tex
\begin{deluxetable}{cccc}
\tabletypesize{\scriptsize}
\tablecaption{TTV Measurements}
\tablewidth{0pt}
\tablehead{
\colhead{$n$}  & \colhead{$t_n$} & \colhead{$TTV_n$}  & \colhead{$TTV_{n\sigma}$} \\
\colhead{}     & \colhead{}      & \colhead{days}     & \colhead{days} 
}
\startdata
KOI-89.01 & & & \\
1 & 83.56887677 & -0.00469193 & 0.00463798 \\  
2 & 168.25615788 & 0.00778280 & 0.00516824 \\  
3 & 252.94343900 & -0.00271064 & 0.00555464 \\  
4 & 337.63072011 & -0.00016954 & 0.00476691 \\  
5 & 422.31800122 & 0.00327206 & 0.00417621 \\  
6 & 507.00528233 & -0.00463520 & 0.00479403 \\  
7 & 591.69256345 & 0.01299045 & 0.00805245 \\
\enddata
\tablecomments{Table 5 is published in its entirety in the electronic edition of the {\it Astrophysical Journal Letters}.  A portion is shown here for guidance regarding its form and content.}
\label{ttvcat}
\end{deluxetable}

%% file: multi.bbl
\begin{thebibliography}{otherstuff}

\bibitem[Adams et al.(2012)]{adams12} Adams, E. R., et al. 2012, \aj, 144 42
\bibitem[Adams et al.(2013)]{adams13} Adams, E. R., et al. 2013, \aj, 146 9
\bibitem[Agol et al.(2013)]{ago13} Agol, E., et al. 2013, in preparation
\bibitem[Akeson et al.(2013)]{ake13} Akeson, R.L., Chen, X., Ciradi, D., et al. 2013, astroph 1307.2944v1
\bibitem[Barclay et al.(2012)]{bar12} Barclay, T., Huber, D., Rowe, J.F., et al. 2013, \apj, 761, 53
\bibitem[Batalha et al.(2010)]{bat10} Batalha, N.M., Rowe, J.F., Gilliland, R.L. et al. 2010, \apj, 713, 103
\bibitem[Batalha et al.(2012)]{bat12} Batalha, N.M., Borucki, W.J., Bryson, S.T. et al. 2011, \apj, 729, 27
\bibitem[Batalha et al.(2013)]{bat13} Batalha, N.M., Rowe, J.F., Bryson, S.T. et al. 2013, ApJS, 204, 24
\bibitem[Borucki et al.(2011)]{bor11} Borucki, W.J., Koch, D.G., Basri, G. et al. 2011, \apj, 736, 19
\bibitem[Borucki et al.(2013)]{Borucki:2013} Borucki, W. J., et al.~2013, Science 340, 587 
\bibitem[Brown et al.(2011)]{bro11} Brown, T.M., Latham, D.W., Everett, M.E., Esquerdo, G.A. 2011, \aj, 142, 112
\bibitem[Bruntt et al.(2012)]{bru12} Bruntt, H., Basu, S., Smalley B. et al. 2012, MNRAS, 423, 122
\bibitem[Bryson et al.(2013)]{bry13a} Bryson, S.T., et al. 2013, PASP, 125, 930
\bibitem[Bryson \& Morton(2013)]{bry13b} Bryson, S.T., Morton, T.D., 2013, in preparation
\bibitem[Buchhave et al.(2012)]{buc12} Buchhave, L.A., Latham, D.W., Johansen, A. et al. 2012, Nature, 486, 375
\bibitem[Burke et al.(2013)]{bur13} Burke, C. et al. 2013, \apj, 210, 19
\bibitem[Carter \& Agol(2013)]{car13} Carter, J.A., Agol, E. 2013, \apj, 765, 132
%\bibitem[Christiansen et al.(2012)]{chi12} Christiansen, J.L., Jenkins, J.M., Barclay T.S., et al. 2012, astroph 1208.0595v1 
\bibitem[Claret \& Bloemen(2011)]{cla11} Claret, A., Bloemen, S. 2011, A\&A, 529, 75
\bibitem[Coughlin et al.(2013)]{cou13} Coughlin, J. et al. 2013 in preparation
\bibitem[Demarque et al.(2004)]{dem04} Demarque, P., Woo, J-H., Kim, Y-C., Yi, S.K. 2004, ApJS, 155, 667
%\bibitem[Dotter et al.(2008)]{dot08} Dotter, A., Chaboyer, B., Jevremovic, D. et al. 2008, ApJS, 178, 89
\bibitem[Ford(2005)]{for05} Ford, E.B. 2005, AJ, 129, 1706
\bibitem[Ford et al.(2011)]{for11} Ford, E.B., Rowe, J.F., Fabrycky, D.C., et al. 2011, ApJS, 197, 2
\bibitem[Fressin et al.(2012)]{fre12} Fressin, F., Torres, G., Rowe, J.F. et al. 2012, Nature, 482, 195
\bibitem[Fressin et al.(2013)]{fre13} Freesin, F., Torres, G., Charbonneau, D., et al. 2013, \apj, 766, 81
\bibitem[Gautier et al.(2012)]{gau12} Gautier, T.N., Charbonneau, D., Rowe. J.F. et al. 2012, \apj, 749, 15
\bibitem[Gregory(2011)]{gre11} Gregory, P.C. 2011, MNRAS, 410, 94
\bibitem[Huber et al.(2013)]{hub13} Huber, D., Chaplin, W.J., Christensen-Dalsgaard, J. 2013, \apj, 767, 127
\bibitem[Jenkins et al.(2002)]{jen02} Jenkins, J.M., Caldwell, D.A., Borucki, W.J. 2002, \apj, 564, 495
\bibitem[Kasting, Whitmire \& Reynolds(1993)]{kas93} Kasting, J.F., Whitmire, D.P., Reynolds, R.T. 1993, Icarus, 101, 108
\bibitem[Kipping(2013)]{kip13} Kipping, D. 2013, astroph 1306.4982v1
\bibitem[Kopparapu et al.(2013)]{kop13} Kopparapu, R.k., Ramirez, R., Kasting J.F. et al. 2013, \apj, 765, 131
\bibitem[Latham et al.(2011)]{lat11} Latham, D.W., Rowe, J.F., Quinn, S.N. et al. 2011, \apj, 732, 24
\bibitem[Lissauer et al.(2011)]{lis11} Lissauer, J.J., Ragozzine, D., Fabrycky, D.C.. et al. 2011, \apjs, 197, 8
\bibitem[Lissauer et al.(2012)]{lis12} Lissauer, J.J., Marcy, G.W., Rowe, J.F. et al. 2012, \apj, 750, 112, Paper I
\bibitem[Lissauer et al.(2013)]{lis13} Lissauer, J.J., Marcy, G.W., Bryson, S.T. et al. 2013, \apj, this issue, Paper II
\bibitem[Mandel \& Agol(2002)]{man02} Mandel, K., Agol, E. 2002, ApJ, 580, 171
%\bibitem[Mann et al.(2012)]{man12} Mann, A.W., Gaidos, E., Lepine, S., Hilton, E.J. 2012, \apj, 753, 90
\bibitem[Marcy et al.(2008)]{mar08} Marcy, G. W., et al. 2008, Physica Scripta, 130, 14001
\bibitem[Marcy et al.(2013)]{mar13} Marcy, G.W., Issacson, H., Howard, A. et al. 2013, \apj, 210, 20
\bibitem[Matijevic et al.(2012)]{mat12} Matijevic, G., Prsa, A., Orosz, J. et al. 2012, \aj, 143, 5
\bibitem[Mazeh et al.(2013)]{maz13} Mazeh, T., Nachmani G., Sokol, G. et al. 2013, A\&A, 541, 56 
\bibitem[More, Garbow \& Hillstrom(1980)]{mor80} More, J.J, Garbow, B.S., Hillstrom, K.E., User Guide for MINPACK-1, Argonne National Laboratory Report ANL-80-74, Argonne, Ill., 1980
\bibitem[Morton \& Johnson(2011)]{mor11} Morton, T.D., Johnson, J.A. 2011, \apj, 738, 170
\bibitem[Petigura et al.(2013)]{pet13} Petigura, E. A., Marcy, G. W. \& Howard, A. 2013, \apj, 770, 69
\bibitem[Plavchan et al.(2012)]{pla12} Plavchan, P., Bilinski, C., Currie, T. 2012, astrph 1203.1887
\bibitem[Pinsonneault et al.(2012)]{pin12} Pinsonneault, M.H., An. D., Molenda-Zakowicz, J. et al. 2012, ApJS, 199, 30
\bibitem[Raghavan et al.(2010)]{rag10} Raghavan, D., McAlister, H.A., Henry, T.J. et al. 2010, ApJS, 190, 1
\bibitem[Rowe et al.(2006)]{row06} Rowe, J.F., Matthews, J.M., Seager, S. et al. 2006, \apj, 646, 1241
\bibitem[Santerne et al.(2013)]{san13} Santerne, A., Fressin, F., Diaz, R. et al. 2013, A\&A, 557, 139
\bibitem[Seager \& Mallen-Ornelas(2003)]{sea03} Seager. S., Mallen-Ornelas 2003, \apj, 585, 1038
%\bibitem[Steffen et al.(2010)]{ste10} Steffen, J.H. et al. 2010, \apj, 725, 1226
\bibitem[Steffen et al.(2012)]{ste12} Steffen, J.H., Fabrycky, D.C., Ford, E.B. et al. 2012, \mnras, 421, 2342
\bibitem[Thompson et al.(2012)]{tho12} Thompson, S.E., Everett, M., Mullally, F. et al. 2012, \apj, 753, 86
\bibitem[Tingley, Bonomo \& Deeg(2011)]{tin11} Tingley, B., Bonomo, A.S., Deeg, H.J., 2011, \apj, 726, 112
\bibitem[Torres et al.(2011)]{tor11} Torres, G., Fressin, F., Batalha, N.M., et al. 2011, \apj, 727, 24
\bibitem[Torres et al.(2012)]{tor12} Torres, G., Fischer, D.A., Sozzetti, A., et al. 2012, \apj, 757, 161
\bibitem[Trimble(1990)]{tri90} Trimble, V. 1990, \mnras, 242, 79
\bibitem[Valenti et al.(1996)]{val96} Valenti, J.A., Piskunov, N. 1996, A\&AS, 118, 595
\bibitem[Walker et al.(2008)]{wal08} Walker, G.A.H., Croll, B., Matthews, J.M. et al. 2008, A\&A, 482, 691 
\bibitem[Weiss et al.(2013)]{wei13} Weiss, L.M., Marcy, G.W., Rowe, J.F. et al. 2013, \apj, 768, 14
\bibitem[Wright et al.(2013)]{wri13}   Wright, T. et al. 2011, PASP, 123, 412
\end{thebibliography}
